\documentclass[a4paper,11pt]{article}
\usepackage[utf8]{inputenc}
\usepackage{lineno}
\usepackage[T1]{fontenc}
\usepackage{amsmath}
\usepackage{amsfonts}
\usepackage{amssymb}
\usepackage{graphicx} %pour afficher des images
\usepackage{hyperref} %pour que les références soient des liens hypertextes
\usepackage{authblk}
\usepackage{cases}
\usepackage{geometry}
\usepackage{natbib}

\usepackage{nameref} % for the crossreference of data availability section

\usepackage{lineno}

\let\oldequation\equation
\let\oldendequation\endequation
\renewenvironment{equation}
  {\linenomathNonumbers\oldequation}
  {\oldendequation\endlinenomath}

\title{Micromechanical controls on the brittle-plastic transition in rocks}
\author{Dong Liu, Nicolas Brantut}
\affil{Department of Earth Sciences, University College London, London, UK} % WC1E6BS,
\date{}

\begin{document}

\maketitle

\begin{abstract}
The rheology of rocks transitions from a localized brittle behaviour to distributed plastic behaviour with increasing pressure and temperature. This brittle-plastic is empirically observed to occur when the material strength becomes lower than the confining stress, which is termed Goetze's criterion. Such a criterion works well for most silicates but is not universal for all materials. We aim to determine the microphysical controls and stress-strain behaviour of rocks in the brittle-plastic transition. We use a micro-mechanical approach due to Horii and Nemat-Nasser, and consider representative volume elements containing sliding wing-cracks and plastic zones. We find solutions for frictional slip, plastic deformation and crack opening at constant confining pressure, and obtain stress-strain evolution. We show that the brittle-plastic transition depends on the confining stress, fracture toughness and plastic yield stress but also critically on the friction coefficient on preexisting defects. Materials with low friction are expected to be more brittle, and experience transition to fully plastic flow at higher pressure than anticipated from Goetze's criterion. The overall success of Goetze's criterion for the brittle-plastic transition in rocks is likely arising from the low toughness, high strength, and medium friction coefficient character of most rock forming minerals. 
\end{abstract}

% \paragraph{Key words:} Defects,  Fracture and flow, Geomechanics, Numerical modelling,  Plasticity, diffusion and creep
% % Keywords:(up to six), Rheology and friction of fault zones, Elasticity and anelasticity, Microstructures
% % reviewer: J.Rudinicki, J. Fortin, H. Bhat, P. Baud?
% % change the template

%\begin{linenumbers} % comment to remove line numbers

\section{Introduction}

The rheology of rocks transitions from brittle to plastic with increasing pressure and temperature. In the laboratory, this transition is marked by the change in deformation structure from pervasive microcracking with localized shear bands to distributed microcracking and homogeneous plastic flow. It is reflected in the stress-strain loading curve as the transition from strain-softening to strain hardening or fully plastic yielding \cite[][chap. 9]{Pate2005}. 

At the microscale, brittle deformation is controlled by frictional sliding on preexisting defects and opening of tensile microcracks; two key physical quantities characterize these processes: friction coefficient and fracture toughness. In addition, geometrical quantities such as grain size or defect orientation and dimension distributions also have a great impact on the brittle behaviour. Plastic flow is controlled by thermally activated intracrystalline processes such as dislocation glide and solid state diffusion, which are related to crystal lattice properties (Burgers vectors, Peierls stress or ``lattice friction'', diffusion coefficients). Experimental rock deformation data show that the switch between brittle and plastic regimes is not abrupt, but that there exists a range of pressure, temperature and strain rate conditions where both brittle and plastic microscale processes coexist \cite[e.g.][]{Rutt1972,FrEv1989,HiTu1994}. In order to determine the rheology of rocks across the brittle-plastic transition, it is crucial to understand how the deformation processes interact and find the combination of physical quantities that would allow us to make reliable extrapolation of laboratory data to the natural scale. 

Based on experimental data \cite{EdPa1972}, Goetze proposed in a private communication \cite{KoEv1995} that fully plastic flow  of rocks (i.e., deformation where cracking is completely suppressed) occurs when the flow stress is about equal to the effective confining pressure. For dry rocks, Goetze's criterion is written
\begin{equation}
    \sigma_1-\sigma_3=\sigma_3
\end{equation}
where $\sigma_1$ is the maximum principal stress and $\sigma_3$ the confining stress. This simple criterion has been shown to be broadly valid in most silicate and carbonate rocks \cite{FrEv1989,FrEv1990,HiTu1994,Drui2011,Ryba2021}, even though it ought to be considered a rule of thumb rather than an exact condition or law. Goetze's criterion reflects the effect of the confining stress in the brittle-plastic transition but the influence of other factors such as temperature, strain rate, pore fluid pressure, grain size, and chemical activities of mineral components is simply lumped into the differential stress, which needs to be measured experimentally. Moreover, it is clear that Goetze's criterion not a universal law applicable to all materials: metals are purely plastic at atmospheric pressure, and talc remains dilatant at pressures that exceed its strength \cite{EdPa1972,Esca2008}. This hints that the apparent validity of Goetze's criterion in most rocks could be due to a similarity in key microphysical properties amongst many rock types. Indeed, most rocks tend to have low toughness \cite[e.g.][]{Atki1984}, and a similar friction coefficient around 0.6 to 0.8 \cite{Byer1978}.

Micromechanical approaches have been developed to investigate the physical controls on the brittle-plastic transition. Nemat-Nasser and his co-workers \cite{NeHo1982,HoNe1985,Nema1983,HoNe1986} used a fracture mechanics model in 2D, dry conditions to predict the onset of microcracking under compression. Their model consists of a thin straight sliding crack with cohesive and frictional resistance, which allows the nucleation of kinked wing cracks and two collinear plastic deformation at the tips under far-field stresses, as shown in Fig.~\ref{fig:StrainRotation}. Using such a model, they explained that the macroscopic faulting or shear failure results from the unstable growth of a row of suitably oriented model flaws, and that a transition from brittle to plastic response takes place due to the gradual suppression of tensile cracks emanating from the tips of a row of flaws with an increasing confining stress. Moreover, this model was the first to include both the influence of the confining pressure and material properties by introducing a ductility parameter $\Delta$ \cite{HoNe1986},
\begin{equation}
    \Delta=\frac{K_\mathrm{Ic}}{\tau_\mathrm{y} (\pi c)^{1/2}},
    \label{eq:ductility}
\end{equation}
where $K_\mathrm{Ic}$ is the mode I fracture toughness, and $\tau_\mathrm{y}$ the yielding stress in shear. $c$ is the half-length of the sliding crack, which characterizes various microscopic inhomogeneities existing in rocks, such as grain boundary cavities, the interface between dissimilar constituents, the intersection of slip bands with an adjacent grain, and other possible material or geometric discontinuities. The influence of temperature or other environmental factors enters implicitly through the associated values of the fracture toughness $K_\mathrm{Ic}$ and yield stress $\tau_\mathrm{y}$. A larger ductility parameter indicates a more ductile behaviour under the same confining stress. 

Interestingly, the definition Eq.~\eqref{eq:ductility} characterizes the brittle-plastic transition in other contexts. In the process of grain size reduction in compression, \cite{Kend1978} showed that a critical particle size exists below which the particles will deform purely plastically and crack propagation becomes impossible; this critical size is given by $32 E G_\mathrm{c}/3\tau_\mathrm{y}^2$, so that the condition for comminution is equivalent to $\Delta \lesssim 1/10$ if we take $c$ as the grain size. In a similar context, \cite{AgMo2016} showed a change in wear mechanism from  fracture-induced debris to plastic deformation when contact junctions fall below a critical length scale proportional to $ E G_\mathrm{c}/\tau_\mathrm{y}^2$, yielding a similar criterion of $\Delta \lesssim 1/3$ for the transition to plastic wear, again using the particle size for $c$ in Eq.~\eqref{eq:ductility}. Finally, the characteristic size obtained by the combination $(K_\mathrm{Ic}/\tau_\mathrm{y})^2$ also scales with (i) the dimension of the crack-tip cohesive zone if we interpret $\tau_\mathrm{y}$ as cohesive stress \cite[e.g.][Eq. 3.16]{Lawn1993}, and (ii) the size of the plastic shielding zone ahead of a crack in a plastic material, if $\tau_\mathrm{y}$ is now interpreted as lattice friction \cite[e.g.][Eq. 7.12]{Lawn1993}. The ductility parameter is thus a key quantity controlling the brittle-plastic transition in materials. 

In the context of the frictional sliding crack model without plasticity, complete stress-strain relations have been derived using a range of kinematic \cite{Kach1982,MoGu1982,NeOb1988} and energy methods \cite{BaGr1998,DeEv2008,Bhat2012}. Recently, \cite{NiFo2017} introduced a coupled brittle-plastic model where deformation results from a superposition of strains due to wing crack growth, bulk plastic flow and microcrack growth due to dislocation pile-ups. In their approach, plastic flow and brittle failure are only coupled through the accumulation of dislocations in grain-boundary pile-ups, and sliding wing cracks are considered independently of this mechanism. To our knowledge, plasticity has not been yet accounted for in a systematical way from a self-consistent kinematic perspective after \cite{HoNe1986}.
The current status of \cite{HoNe1986}'s results makes the interpretation of experimental data rather challenging: (i) the assumption of a constant ratio between the confining stress and the maximum stress is not consistent with commonly applied loading conditions in the laboratory; (ii) Limited configurations are given by the modelling results, and running the numerical simulation for other configurations is not straightforward (iii) Stress-strain relations which can be directly linked to the measurement in the laboratory have not been derived.

Our work here aims to revisit the frictional sliding crack model in \cite{HoNe1986} accounting for the growth of wing tensile cracks and plastic zones, by using an efficient algorithm based on the Gauss-Chebyshev quadrature and barycentric interpolation techniques. We perform a dimensional analysis and try to relate the physical parameters entering the model description (friction coefficient, toughness, yield stress) to experimentally measurable quantities such as differential stress, axial and volumetric strains, and establish how the model predictions fit the empirical criterion attributed to Goetze. We provide here a tunable numerical tool to investigate the effect of different parameters on the brittle-plastic transition in rocks (see \nameref{sec:data} for more details).

\section{Frictional sliding crack model}
\subsection{Problem formulation}

\begin{figure}
    \centering
    \includegraphics[width=0.6 \linewidth]{ 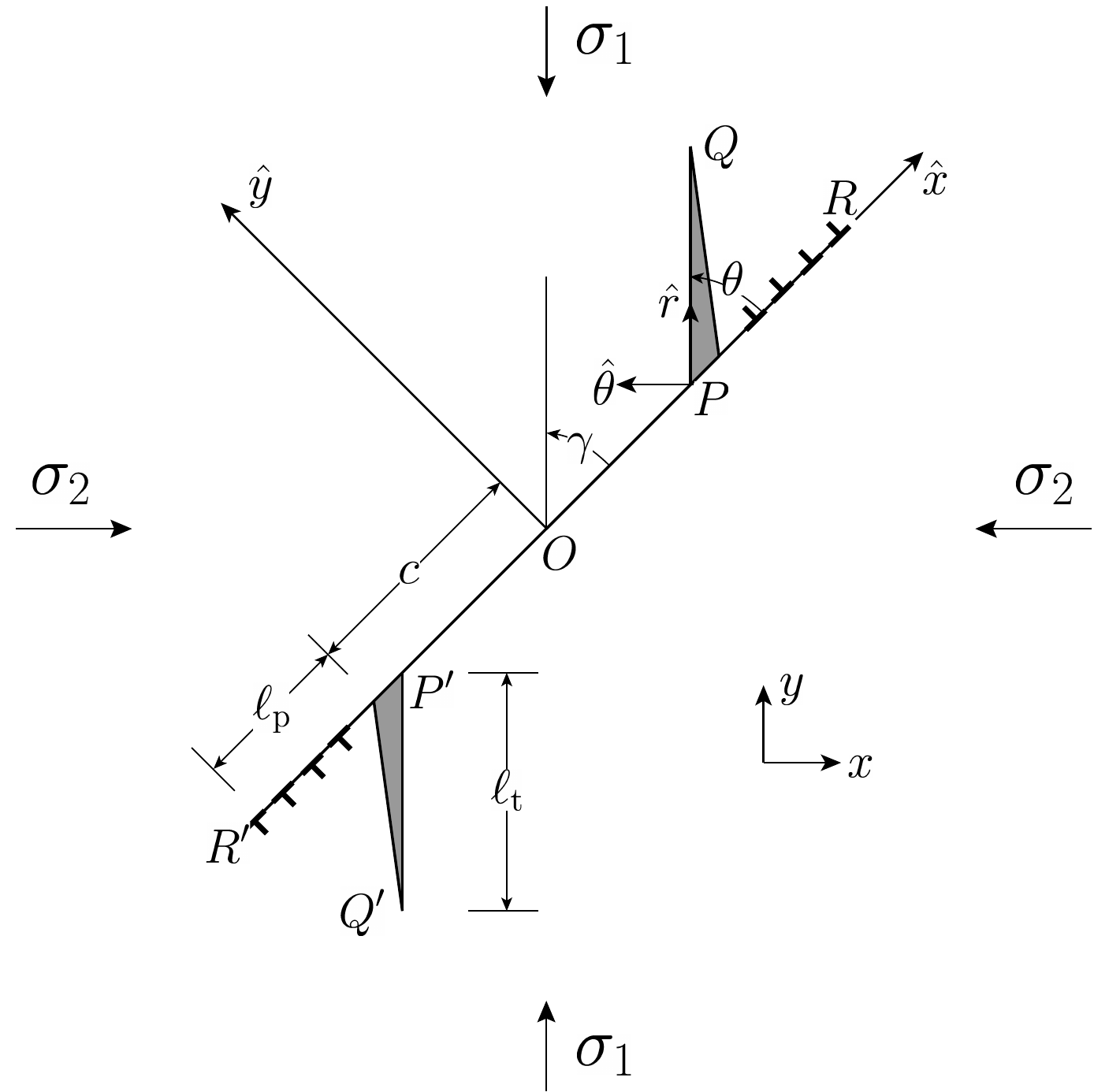}
    \caption{Illustration of the frictional sliding crack model: the frictional sliding crack $PP^{\prime}$, its associated collinear plastic zones $PR$ and $P^{\prime}R^{\prime}$, the kinked wing cracks $PQ$ and $P^{\prime}Q^{\prime}$ and their corresponding local coordinate system: $\hat{x}, \hat{y}$ for the preexisting shear crack and the plastic zones, $\hat{r}, \hat{\theta}$ for the wing cracks.  \cite{HoNe1986}.}
    \label{fig:StrainRotation}
\end{figure} 

In the following, we focus on a simplified model with an isolated frictional sliding crack with the possible presence of straight tensile wing cracks and collinear plastic zones at the tips. This 2D micromechanical model was first proposed by \cite{NeHo1982,HoNe1985,Nema1983,HoNe1986}, and we refer the reader to these sources for a detailed presentation.

Fig.~\ref{fig:StrainRotation} illustrates the geometry under consideration, which consists of a pre-existing shear flaw $PP^{\prime}$ with the size of $2c$, two wing cracks $PQ$ and $P^{\prime}Q^{\prime}$ with the length of $\ell_{\mathrm{t}}$, and two collinear plastic zones $PR$ and $P^{\prime}R^{\prime}$ with the length of $\ell_{\mathrm{p}}$. Under axial $\sigma_1$ and lateral compression $\sigma_2$ in the far field ($\sigma_1>\sigma_2$, and compression is viewed positive), the preexisting shear crack slides along a plane with an inclination $\gamma$ with respect to the maximum compression $\sigma_1$. The two wing cracks emanate from the ends of the sliding crack with an inclination $\theta$.

The pre-existing straight crack $PP^{\prime}$ is assumed to remain closed during the entire crack extension process. With a coordinate system $(\hat{x}, \hat{y})$ as shown in Fig.~\ref{fig:StrainRotation}, the shear stress along the pre-existing flaw writes
\begin{equation}
\tau_{\hat{x}\hat{y}}=\tau_{\mathrm{c}}+\mu \sigma_{\hat{y}},
\end{equation}
where $\mu$ is the friction coefficient, $\tau_{\mathrm{c}}$ is the material cohesion,  and $\sigma_{\hat{y}}$ is the normal stress applied on the preexisting flaw $PP^\prime$. 

The wing cracks grow from the ends of the pre-existing flaw which imposes a zero dislocation $[v]$ normal to the slip line at the wing crack tips $P$, $P^{\prime}$:
\begin{equation}
    [v_{\hat{y}}]=0~\text{at}~P~\text{and} ~P^{\prime}.
\label{eq:boundarywing}
\end{equation}
Since $PQ$ and $P^{\prime}Q^{\prime}$ are tensile cracks, the wing crack surfaces are stress-free in dry conditions:
\begin{equation}
    \sigma_{\hat{\theta}}=\tau_{\hat{r} \hat{\theta}}=0,
    \label{eq:governingwing}
\end{equation}
where $\hat{r}, \hat{\theta}$ indicate a local coordinate system along the wing crack surfaces which can be obtained by rotating $\hat{x}$, $\hat{y}$ with an angle of $\theta$ as shown in Fig.~\ref{fig:StrainRotation}.

The plastic zones are modelled by dislocation arrays collinear with the pre-existing flaw. The stresses along $PR$ and $P^{\prime}R^{\prime}$ are bounded with the constant yielding shear stress $\tau_{\mathrm{y}}$:
\begin{equation}
    \tau_{\hat{x}\hat{y}}=\tau_{\mathrm{y}}.
    \label{eq:governingplastic}
\end{equation}
As a result, the mode II stress intensity factor at the ends of the plastic zones must be zero. Following \cite{HoNe1986}, we set:
\begin{equation}
    K_{\mathrm{II}}^R=0,\quad \text{at }R \text{ and } R^{\prime}.
    \label{eq:boundaryplastic}
\end{equation}
The boundary conditions along the frictional sliding flaw, open wings and plastic zones, together with elastic equilibrium, constitute a boundary value problem in the form of singular integral equations. The complete, exact formulation of the problem is given in Appendix.~\ref{ax:Formulation}. We rewrite the equations in a dimensionless form by scaling length quantities with the half flaw size $c$ and stress quantities with the yield shear stress $\tau_{\mathrm{y}}$. Assuming a constant confining stress $\sigma_2/\tau_{\mathrm{y}}$ with a well-defined geometry (given values of $\gamma$, $\theta$, and assumed values of $\ell_{\mathrm{t}}/c$, $\ell_{\mathrm{p}}/c$), we solve the singular integral equations numerically for the maximum stress $\sigma_1/\tau_{\mathrm{y}}$ and the dislocation densities along the wing cracks $\alpha_\mathrm{t}$ and the plastic zones $\alpha_\mathrm{p}$. Based on these dislocation densities, we obtain the crack opening or shear displacement $[v]$ along the crack surfaces, and evaluate the stress intensity factors $K_{\mathrm{I}}/(\tau_{\mathrm{y}}\sqrt{\pi c})$, $K_{\mathrm{II}}/(\tau_{\mathrm{y}}\sqrt{\pi c})$ at the crack tips. 

\subsection{Micro-strain calculation}
We estimate the deformation due to the preexisting flaw and its associated wing cracks and plastic zones by considering a representative elementary volume around the frictional sliding crack. Neglecting interactions between possibly neighbouring flaws (i.e., considering the representative elementary volume as isolated) and viewing contraction positive, the total micro-strain consists of an elastic strain due to the elastic deformation of the matrix $\epsilon^{\mathrm{e}}$, plus an inelastic strain $\epsilon^{\mathrm{p}}$ due to the sliding of the pre-existing flaw, the dislocations in the plastic zones and tensile crack opening of the wings: $\epsilon^{\mathrm{t}}=\epsilon^{\mathrm{e}}+\epsilon^{\mathrm{p}}$. Using the hydrostatic condition ($\sigma_1=\sigma_2$) as our reference configuration, $\epsilon^{\mathrm{e}}$ characterizes the elastic deformation increment due to the increase of a differential stress of $(\sigma_1-\sigma_2)$. Under plane strain conditions, we have
\begin{equation}
    \epsilon_{xx}^{\mathrm{e}}=-\frac{\nu}{1-\nu}\frac{\sigma_1-\sigma_2}{E^{\prime}}, \quad \epsilon_{yy}^{\mathrm{e}}=\frac{\sigma_1-\sigma_2}{E^{\prime}}
, \quad
    \epsilon_{v}^{\mathrm{e}}=\epsilon_{xx}^{\mathrm{e}}+\epsilon_{yy}^{\mathrm{e}}=\frac{1-2\nu}{1-\nu} \frac{\sigma_1-\sigma_2}{E^{\prime}},    
\end{equation}
where $E^{\prime}=E/(1-\nu^2)$ indicates the plane-strain elastic modulus and $\nu$ Poisson's ratio.

We calculate the inelastic strains by integrating the dislocations on crack surfaces \cite{Kach1982,NeOb1988}, which we obtain by solving the governing equations for the sliding crack model. The total inelastic strain results from the sum of micro-strains associated with all the flaws present in the material:
\begin{equation}
    \epsilon_{ij}^{\mathrm{p}}= \frac{f}{2c^2}\int_{\ell} \{n_i [v_j]+n_j[v_i]\} \text{d}x, \quad x\in \ell,
    \label{eq:inelasticstrain}
\end{equation}
where $f$ represents the density of the flaws of length $2c$, $\mathbf{n}$ is a unit vector normal to a crack surface $\ell$, and $[v_i]$ is the $i^{\mathrm{th}}$ component of displacement on the crack surface. We refer to Appendix.~\ref{ax:Strain} for detailed expressions of the inelastic strains.

The inelastic deformation $\epsilon^\mathrm{p}$ depends on material properties and loading conditions. In the following, we define a normalized inelastic strain $(E^{\prime}/\tau_{\mathrm{y}})\epsilon^{\mathrm{p}}/f$ and a normalized differential stress $(\sigma_1-\sigma_2)/\tau_{\mathrm{y}}$ independent of the material properties.

\section{Solution method}

We adopt a numerical method combining the Gauss-Chebyshev quadrature and barycentric interpolation techniques \cite{ViGa2017}, which allows for a spectral accuracy when solving the singular integral equations that appear in various fracture problems \cite{Gara2019,LiLG2019}. In this study, we solve for half of the problem due to its symmetrical nature: we only consider the dislocations along on one wing crack and one plastic zone. We refer to Appendices.~\ref{ax:Numerical} \ref{ax:Discretization} for detailed discretization of the governing equations.

In the following, we present the sequence of modelling steps used to determine the stress-strain behaviour under constant applied confining stress.

(1) We set up the configuration by imposing constant values to $\gamma$, $\theta$, $\mu$, and $\tau_{\mathrm{c}}/\tau_{\mathrm{y}}$ for different loading conditions $\sigma_2/\tau_{\mathrm{y}}$. 

\begin{figure}
    \centering
    \begin{tabular}{cc}
    \includegraphics[width=0.45 \linewidth]{ 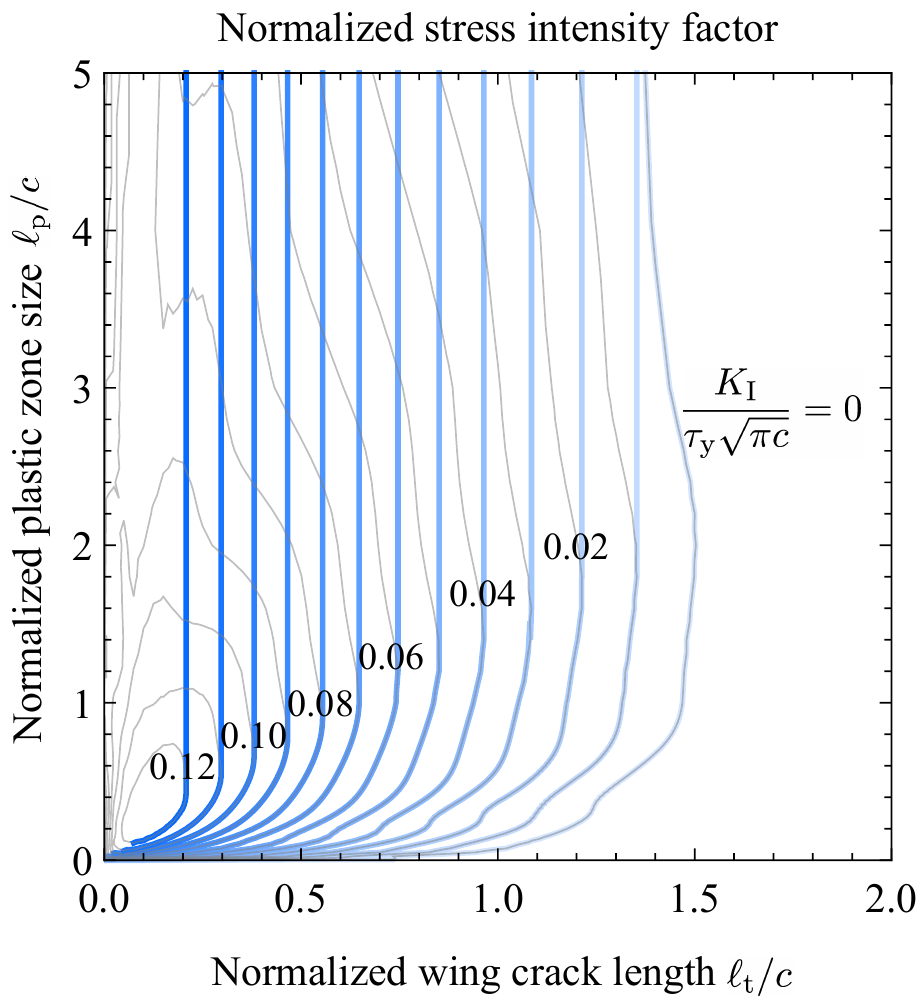}&
    \includegraphics[width=0.45 \linewidth]{ 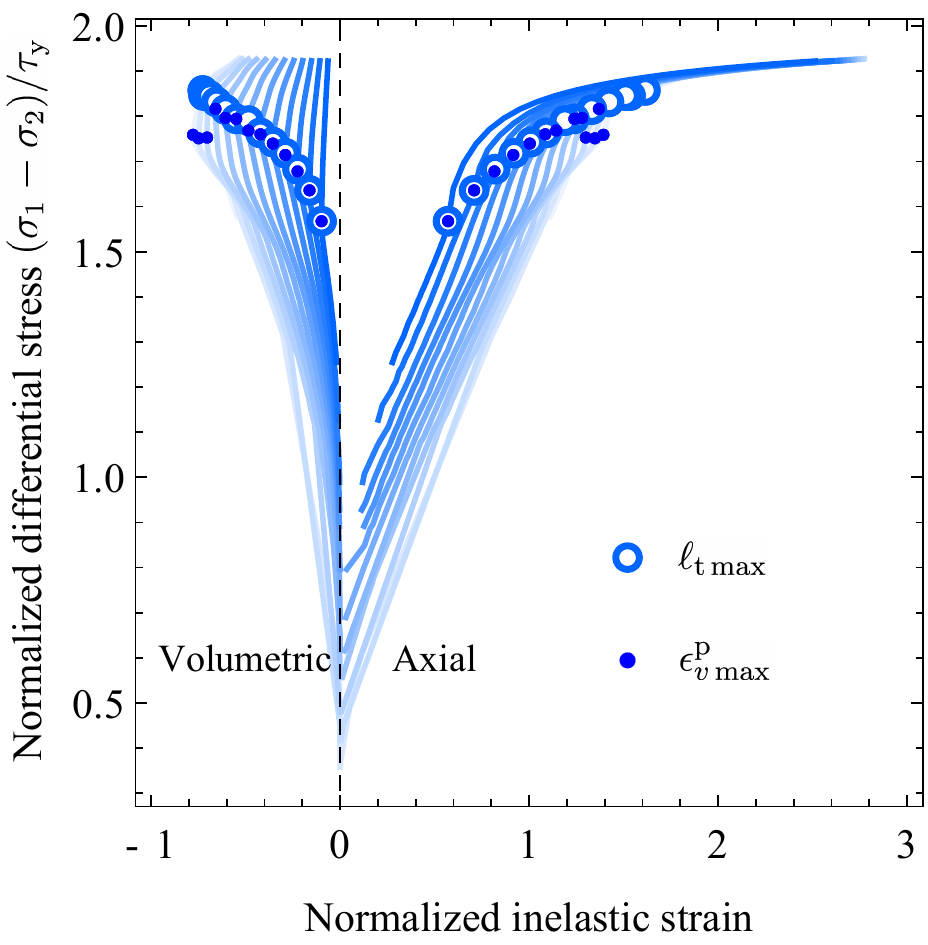}\\
    (a) & (b)
    \end{tabular}
    \caption{(a) Contours (in gray) of normalized stress intensity factor at the wing crack tip $K_{\mathrm{I}}/(\tau_{\mathrm{y}} (\pi c)^{1/2})$ and the solution curves (in blue) for different ductility parameters ($\gamma=\theta=\pi/4$, $\mu=0.6$, $\tau_{\mathrm{c}}=0$ and $\sigma_2/\tau_\mathrm{y}=0.1$).(b) Relation between the normalized differential stress $(\sigma_1-\sigma_2)/\tau_\mathrm{y}$ and the inelastic axial strain $(E^{\prime}/\tau_{\mathrm{y}})\epsilon_{yy}^{\mathrm{p}}/f$, and the inelastic volumetric strain $(E^{\prime}/\tau_{\mathrm{y}})\epsilon_{v}^{\mathrm{p}}/f$. The blue colors with different saturation degrees correspond to the results of different ductility parameters as indicated in panel a). The empty and filled circles in panel b) correspond to the time at which the tensile crack length $\ell_\mathrm{t}$ (empty), and the absolute value of the volumetric strain $\epsilon_v^\mathrm{p}$(filled) reaches their maximum.}
    \label{fig:Illustration}
\end{figure}

(2) For a given $\sigma_2/\tau_{\mathrm{y}}$, we calculate the normalized stress intensity factors $K_{\mathrm{I}}/(\tau_{\mathrm{y}}\sqrt{\pi c})$ and $K_{\mathrm{II}}/(\tau_{\mathrm{y}}\sqrt{\pi c})$, and the normalized axial stress $\sigma_1/\tau_{\mathrm{y}}$ as functions of the normalized wing crack length $\ell_{\mathrm{t}}/c$ and the normalized plastic zone size $\ell_{\mathrm{p}}/c$. Typical results are shown in Fig.~\ref{fig:Illustration} where contours of the mode-I stress intensity factor $K_{\mathrm{I}}/(\tau_{\mathrm{y}}\sqrt{\pi c})$ at the wing crack tip $Q$ are plotted in the $\ell_{\mathrm{t}}/c$, $\ell_{\mathrm{p}}/c$-plane.

(3) We then extract a solution curve in the $\ell_{\mathrm{t}}/c$, $\ell_{\mathrm{p}}/c$-plane by following the criteria in \cite{HoNe1986}:
\begin{equation}
\begin{cases}
     K_{\mathrm{I}}=K_{\mathrm{Ic}},\, \left(K_{\mathrm{I}}/(\tau_{\mathrm{y}} \sqrt{\pi c})=\Delta\right), \text{when tensile wing cracks $\ell_{\mathrm{t}}/c$ are growing},\\
     0<K_{\mathrm{I}}<K_{\mathrm{Ic}},\, \left( 0<K_{\mathrm{I}}/(\tau_{\mathrm{y}} \sqrt{\pi c})<\Delta\right), \text{when tensile wing cracks $\ell_{\mathrm{t}}/c$ are stationary},\\
     K_{\mathrm{I}}=0, \, \left(K_{\mathrm{I}}/(\tau_{\mathrm{y}} \sqrt{\pi c})=0\right), \text{when tensile wing cracks $\ell_{\mathrm{t}}/c$ are closing}, 
\end{cases}
\end{equation}
where $K_{\mathrm{Ic}}$ is the prescribed fracture toughness and  $\Delta=K_{\mathrm{Ic}}/(\tau_{\mathrm{y}} \sqrt{\pi c})$ its normalized form defined as the ductility parameter.
We show an example of the solution curves obtained using such criteria in Fig.~\ref{fig:Illustration}a. The solution curve always initiates at a location closest to the origin of $\ell_{\mathrm{t}}/c$, $\ell_{\mathrm{p}}/c$-plot. %the growth of the wing crack and the plastic zone initiates at a location closest to the origin of $\ell_{\mathrm{t}}/c$, $\ell_{\mathrm{p}}/c$-plot by following the solution curve. % and follows the solution curve under constant confining stress.  

It is interesting to notice that there might be two contours of stress intensity factors sharing the same value, one indicating a more brittle behavior (with a solution curve similar to those shown in Fig.~\ref{fig:Illustration}) and the other a more plastic behavior (with a very large $\ell_{\mathrm{p}}/c$ developed before the nucleation of the wing cracks). This has been reported in \cite{HoNe1986} as a transitional regime where transitions between the two contours in the $\ell_{\mathrm{t}}/c$, $\ell_{\mathrm{p}}/c$-plane may happen during the loading process. In this study, we extract the solution curve using the contour with a larger value of $\ell_{\mathrm{t}}/c$. This leads to a maximizationn of the brittle behaviour.

(4) We re-run the numerical simulations using combinations of $\ell_{\mathrm{t}}/c$ and $\ell_{\mathrm{p}}/c$ extracted from the solution curves for different ductility parameters $\Delta$. Based on the obtained dislocation densities on the crack surfaces, we estimate the corresponding evolution of the normalized differential stress and strain as illustrated in Fig.~\ref{fig:Illustration}b. 

\section{Results}

In the following, we focus on the case of $\gamma=\theta=\pi/4$. Such a geometry is chosen based on two reasons: (i) $\gamma=\pi/4$ maximizes the shear stress on the sliding crack under the bi-axial compression $\tau_{\mathrm{max}}=(\sigma_1-\sigma_2)/2$, thus promoting the crack sliding and deformation, (ii) $\theta=\gamma$ corresponds a direction where wing cracks propagate perpendicular to the confining stress, which is the correct asymptotic behavior for long wings: in a more elaborate model of wing crack growth allowing for orientation change \cite{HoNe1986}, tensile cracks initially appear at $\theta\approx 70^\circ$ and then shortly transition to a direction perpendicular to the confining stress with $\theta=\gamma$. 

With a $45^\circ$-inclined preexisting flaw and straight wing cracks perpendicular to the confining stress, we study the effect of imposed confining stress, friction coefficient and material cohesion using the following values:
\begin{itemize}
\item $\tau_{\mathrm{c}}/\tau_{\mathrm{y}}=0$, $\mu= 0,\, 0.1,\, 0.3,\, 0.6,\, 0.8$. 
\item $\tau_{\mathrm{c}}/\tau_{\mathrm{y}}=0.5$, $\mu=0$.
\end{itemize}
Note that $\mu=0.6$ indicates a friction coefficient representative for most silicates, $\mu=0.1$ represents a low friction coefficient in materials like talc presenting very weak slip planes. The case of zero friction and nonzero cohesive stress is representative of slip on the initial flaw driven by dislocation glide (viewing the flaw as a slip band), where the extended plastic zones with stress $\tau_\mathrm{y}>\tau_\mathrm{c}$ are neighbouring slip bands with higher lattice friction. 

All the results reported in this study are obtained by discretizing the wing crack and the plastic zone using the same number of nodes $n=25$. 
%When extracting the solution curves, we do not set constraints on the growth of the wing crack and plastic zone.
The numerical simulations allow the growth of wing cracks up to $\ell_{\mathrm{t}}/c \sim 10$ (see Appendix.~\ref{ax:Results} for more details) and that of plastic zones with a maximum size of $\ell_{\mathrm{p}}/c \sim 5-12$. 

\subsection{Evolution of the solution curves}
Under constant confining stress, the growth of wing cracks and plastic zones competes with each other across the brittle-plastic transition: Propagation of wing cracks $\ell_{\mathrm{t}}/c$ induces %mode I and mode II
micro-cracking together with dilatant deformation favouring a more brittle behavior, while extension of the plastic yielding zone $\ell_{\mathrm{p}}/c$ corresponds to a more ductile behavior by accumulating more dislocations with little volumetric dilantancy. 

As shown in Fig.~\ref{fig:Illustration}, for materials characterized by small ductility index $\Delta$, as we increase the differential stress while keeping a constant confining stress, the slip along the preexisting shear crack increases,  accompanied by the nucleation of a yielding plastic zone and an increase of the stress intensity factors at the tips. When such a stress-intensity is sufficiently large to fulfil the wing crack nucleation criterion $K_{\mathrm{I}}=K_{\mathrm{Ic}}$, the wing cracks start to grow monotonically together with the plastic zone, and result in an increase of the volumetric deformation. However, the wing cracks can not propagate to arbitrarily large dimension since the increase of the wing crack length $\ell_{\mathrm{t}}/c$ will relax the stress concentration at the tips %and compensate the increase of the associated stress intensity factors due to 
under lateral compression. As a result, there exists a maximum wing crack size $\ell_\mathrm{t\, max}/c$ beyond which the stress intensity factor starts to drop below the material toughness $K_{\mathrm{Ic}}$. Notably, this $\ell_{\mathrm{t\, max}}/c$ often corresponds to the maximum volumetric deformation.  %However, the total volumetric deformation keeps rising when the decrease of the volumetric strain due to the gradual closure of wing cracks is overcompensated by the increasing plastic deformation (Fig.~\ref{fig:Illustration}). 
As the differential stress continues to increase, the tensile crack stops at $\ell_{\mathrm{t}}=\ell_{\mathrm{t\, max}}$ and starts closing with a smaller opening and decreasing stress intensity factors at the tips ($0<K_{\mathrm{I}}<K_{\mathrm{Ic}}$). The increase of the load is then mostly associated with plastic deformation together with an increasing plastic zone $\ell_{\mathrm{p}}/c$. A decrease of the volumetric strain may occur at this stage due to the gradual closure of wing cracks. When the stress intensity factors drop to zero $K_{\mathrm{I}}=0$,  the growth of $\ell_{\mathrm{t}}/c$ and $\ell_{\mathrm{p}}/c$ will follow the contour of $K_{\mathrm{I}}=0$, along which the wing cracks may be partially closed. 

% some kind of discussion here related to the model
We show in Fig.~\ref{fig:Illustration}b a series of illustrative stress-strain curves obtained from the extracted solution curves. Interestingly, the tensile cracking starts quite early, even when the differential stress is low. This implies that the macroscopic inelastic deformation occurs even when the differential stress is much smaller than the plastic yielding limit.
The inelastic strains increase monotonically with the differential stress and all converge to a maximum value when the maximum shear stress is equivalent to the maximum shear strength under the bi-axial loading $\tau_{\mathrm{max}}=(\sigma_1-\sigma_2)/2 \approx \tau_{\mathrm{y}}$.

\subsection{Influencing factors of the brittle-plastic transition}

\paragraph{Effect of the ductility parameter}
The growth of the wing crack and plastic zone is closely related to the ductility parameter. For a given flaw size, an increase of the fracture toughness $K_{\mathrm{Ic}}$, or a decrease of the yielding strength in shear $\tau_{\mathrm{y}}$ may both lead to the increase of the ductility. The former limits the growth of the wing cracks and lowers the volumetric deformation, while the latter favours the yielding of the plastic zone and leads to a larger differential stress with less volumetric strains. As illustrated in Fig.~\ref{fig:Illustration}, the maximum of the wing crack size $\ell_{\mathrm{t}}/c$ decreases with the increase of the ductility. 

When the ductility is sufficiently large, the growth of plastic zones is dominant over that of wing cracks: plastic zones can develop to a much larger size before the nucleation of the wing cracks.  The maximum volumetric strain may depend on the maximum value of $\ell_{\mathrm{p}}$ allowed in the numerical simulations rather than the maximum value of the wing cracks $\ell_{\mathrm{t\, max}}$.
When $\Delta=\infty$,  the deformation is fully plastic with no nucleation of the wing cracks or any volumetric deformation. 

\begin{figure}
    \centering
    \begin{tabular}{cc}
    \includegraphics[width=0.45 \linewidth]{ 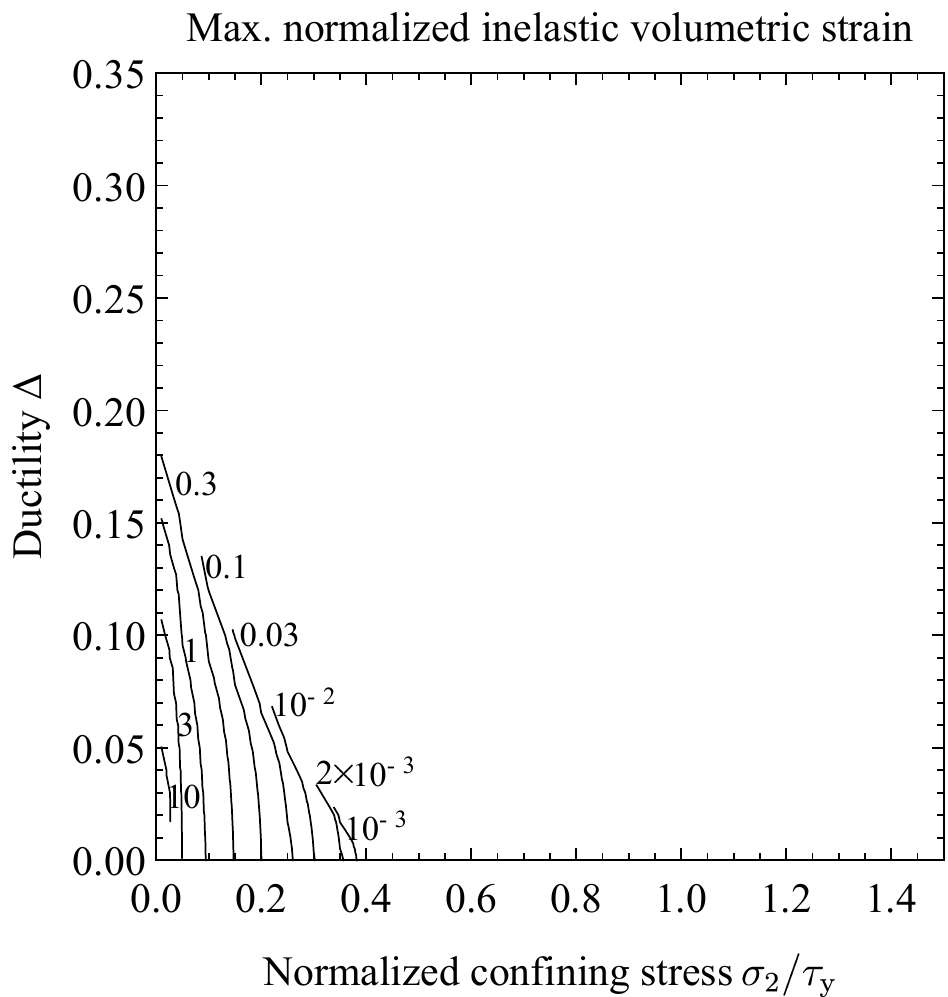}&
    \includegraphics[width=0.45 \linewidth]{ 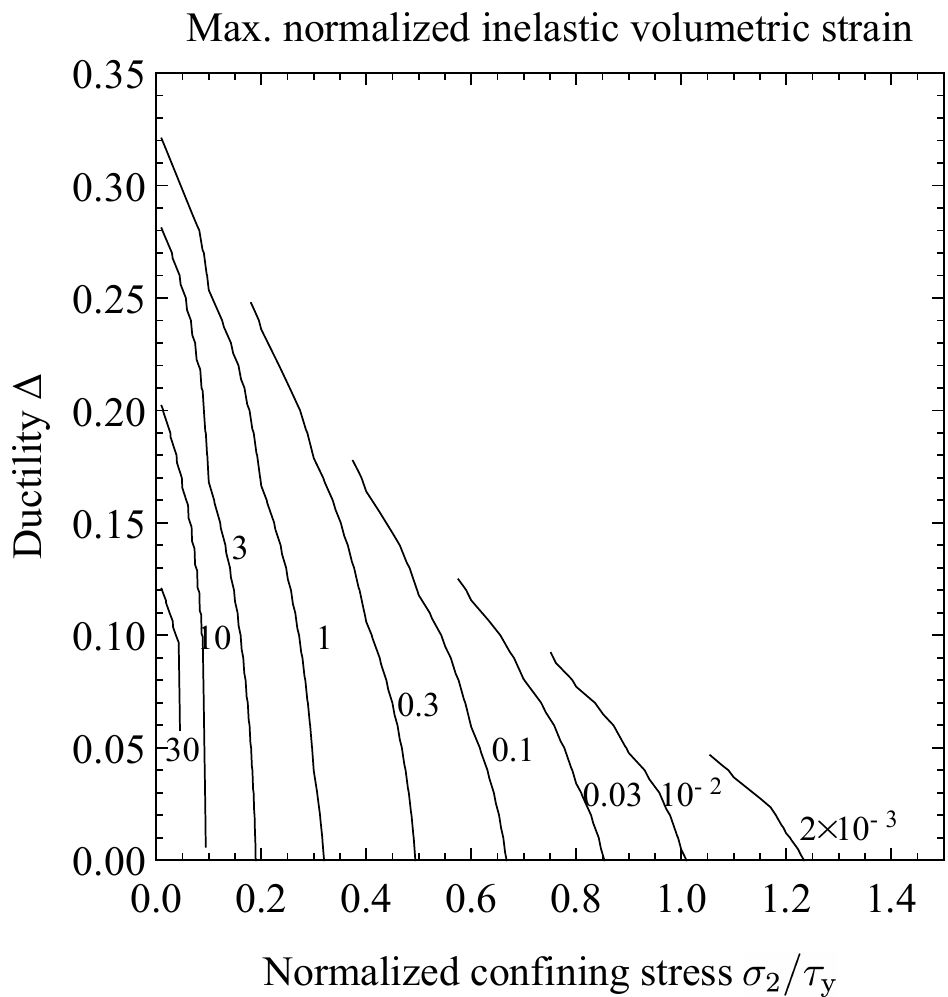}\\
    (a) $\mu=0.6$ & (b) $\mu=0.1$
    \end{tabular}
    \caption{Contours of constant normalized volumetric strains $\epsilon_{v\, \mathrm{max}}^{\mathrm{p}} (E^{\prime}/\tau_{\mathrm{y}})/f$ for the cases of $\gamma=\theta=\pi/4$ and $\tau_{\mathrm{c}}=0$ with different friction coefficients.
    %: (a) $\mu=0.6$ and  (b) $\mu=0.1$.
    }
    \label{fig:EvContour}
\end{figure}

\paragraph{Effect of confining stress}
An increased confining stress which acts perpendicular to the wing cracks tends to limit the growth of wing cracks and reduce their opening. The volumetric strain, which mainly results from the opening of the wing cracks, thus decreases with increasing confining pressure (as shown in Fig.~\ref{fig:EvContour}). In other words, a large confining stress favours the growth of the plastic zone, and tends to result in more plastic deformation than cracking. Materials presenting a smaller value of ductility (more volumetric deformation/ more brittle cracking) are more sensitive to the change of confining stress.

\paragraph{Effect of friction coefficient}
A smaller friction coefficient $\mu$ favours the sliding of the preexisting flaw, and thus the growth of wing cracks. This leads to more volumetric deformation and thus a more brittle behaviour. 
As shown in Fig.~\ref{fig:EvContour}, the confining stress necessary to reach the  fully plastic regime (near zero volumetric strain) decreases with an increasing friction coefficient.  %if we define the brittle-plastic transition occurs at $\epsilon_{v, \mathrm{max}}^{\mathrm{p}}(E^{\prime}/\tau_{\mathrm{y}})/f=10^{-2},\, 0.1,\, 1$,

\begin{figure}
    \centering
    \begin{tabular}{cc}
    \includegraphics[height=0.45 \linewidth]{ 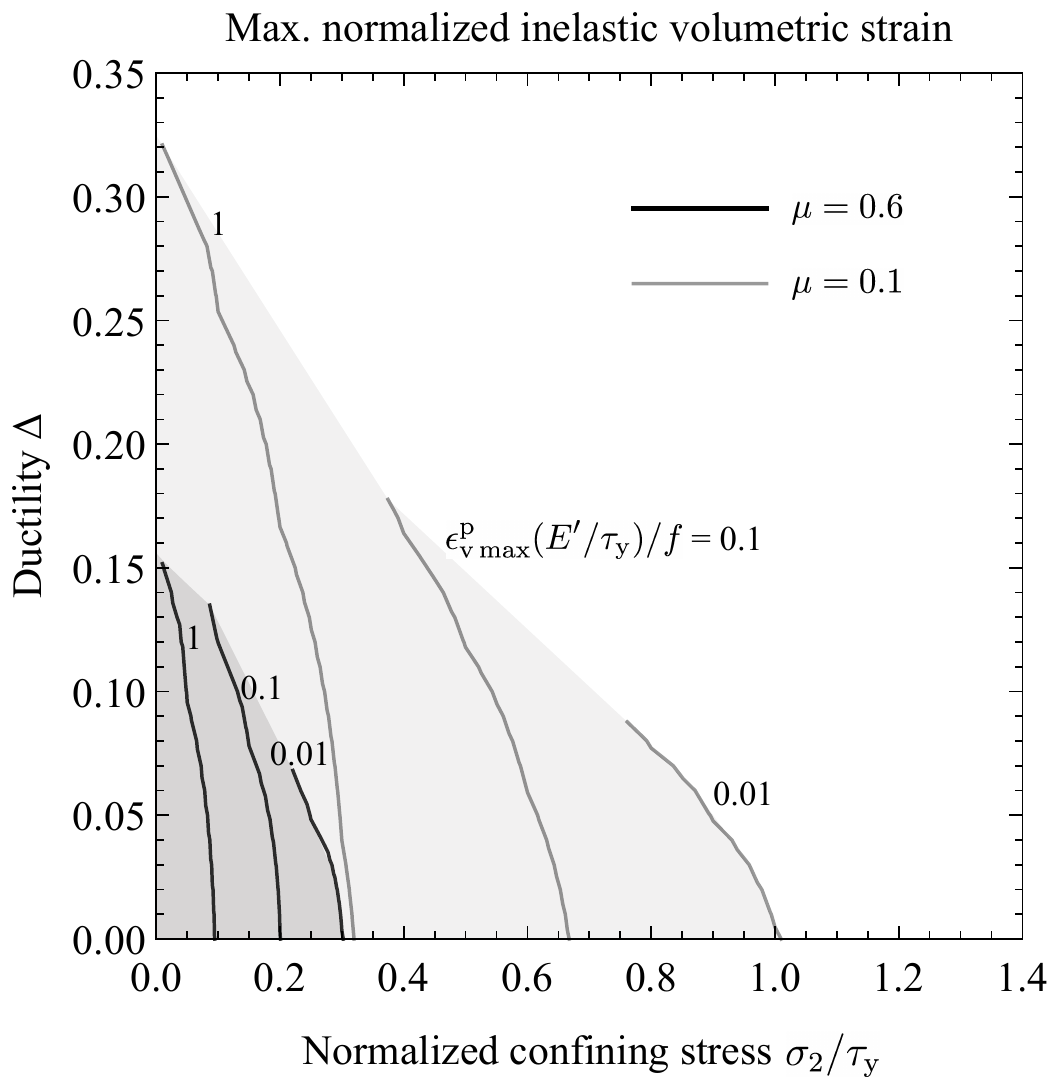}&
    \includegraphics[height=0.45 \linewidth]{ 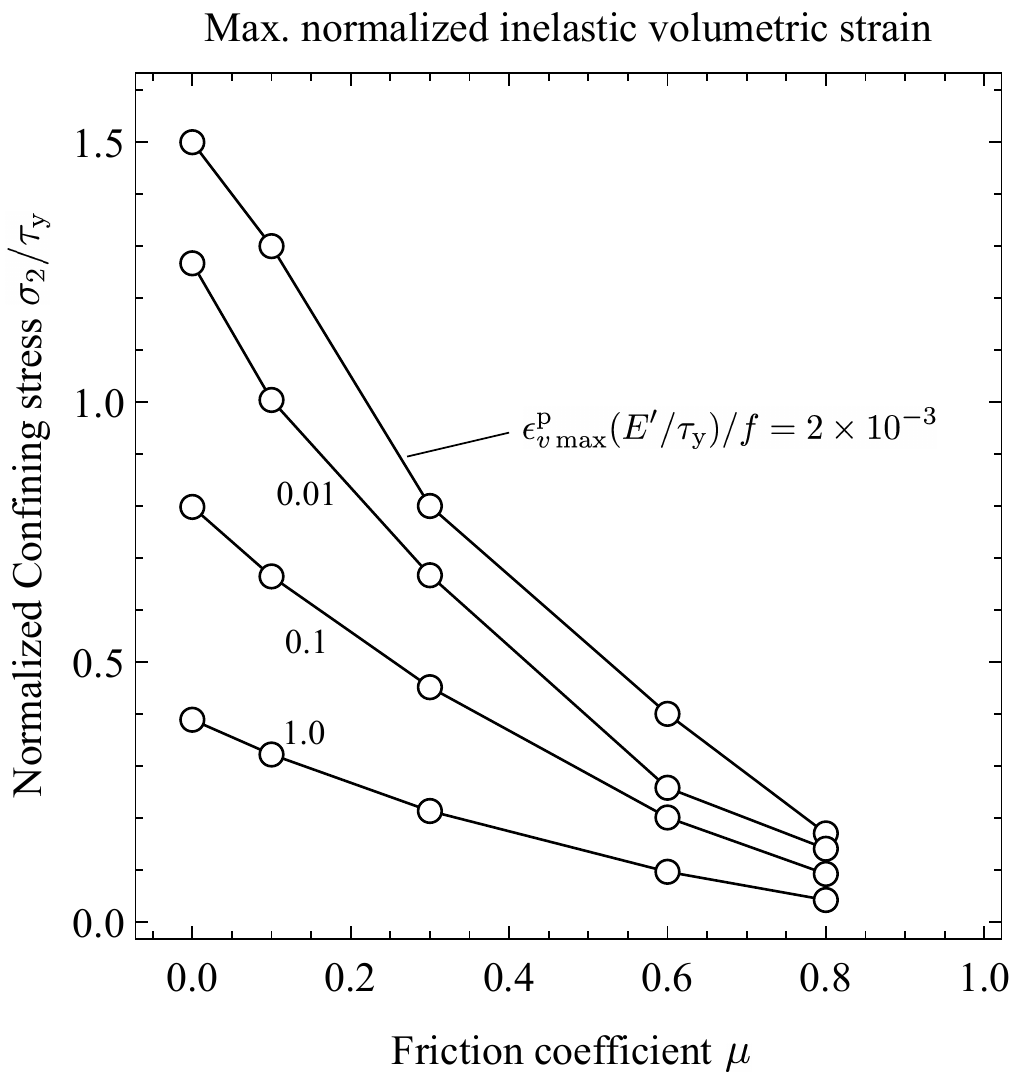}\\
    (a) & (b)
    \end{tabular}
    \caption{%Effect of friction coefficient. (a) Contours of  normalized volumetric strain $\epsilon_{v\,\mathrm{max}}^{\mathrm{p}} (E^{\prime}/\tau_{\mathrm{y}})/f$ for $\mu=0,\, 0.1,\, 0.3,\, 0.6,\, 0.8$ ($\gamma=\theta=\pi/4,\,\tau_{\mathrm{c}}=0$). A darker gray indicates a larger friction coefficient.  (b) Confining stress necessary at $\Delta=0$ to obtain different volumetric deformation $\epsilon_{v\,\mathrm{max}}^{\mathrm{p}}(E^{\prime}/\tau_{\mathrm{y}})/f$ as a function of friction coefficient.
    Effect of friction coefficient for $\gamma=\theta=\pi/4,\,\tau_{\mathrm{c}}=0$. (a) Contours of  normalized volumetric strain $\epsilon_{v\,\mathrm{max}}^{\mathrm{p}} (E^{\prime}/\tau_{\mathrm{y}})/f$ for $\mu=0.1,\, 0.6$.  (b) Maximum confining stress (at $\Delta=0$) necessary to obtain certain volumetric deformation $\epsilon_{v\,\mathrm{max}}^{\mathrm{p}}(E^{\prime}/\tau_{\mathrm{y}})/f$ for different friction coefficients.
    }
    \label{fig:MuEffect}
\end{figure}

The dependency on ductility becomes negligible when the confining stress is sufficiently large. As shown in Fig.~\ref{fig:EvContour}, little volumetric strain occurs for $\mu=0.6$ regardless of the ductility when the confining stress $\sigma_2>0.4 \tau_{\mathrm{y}}$. As shown in Fig.~\ref{fig:MuEffect}b,  we collect the confining stresses at $\Delta=0$ with different volumetric deformation. They indicate the upper limit of the confining stress necessary to remove any cracking/dilatant deformation. Increasing the confining stress is seen to have a more significant effect, i.e., promote an earlier brittle-plastic transition, for materials with a larger friction coefficient (Fig.~\ref{fig:MuEffect}). 

\begin{figure}
    \centering
    \begin{tabular}{cc}
    \includegraphics[width=0.45 \linewidth]{ 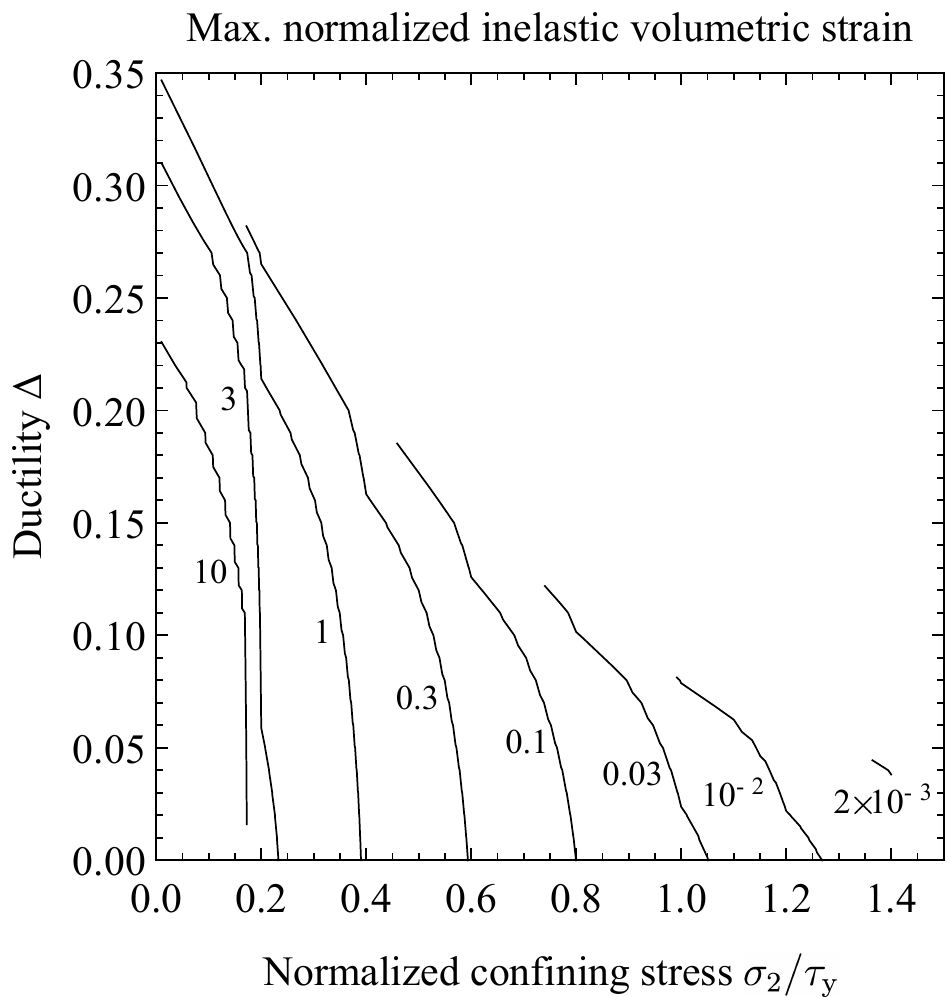}&
    \includegraphics[width=0.45 \linewidth]{ 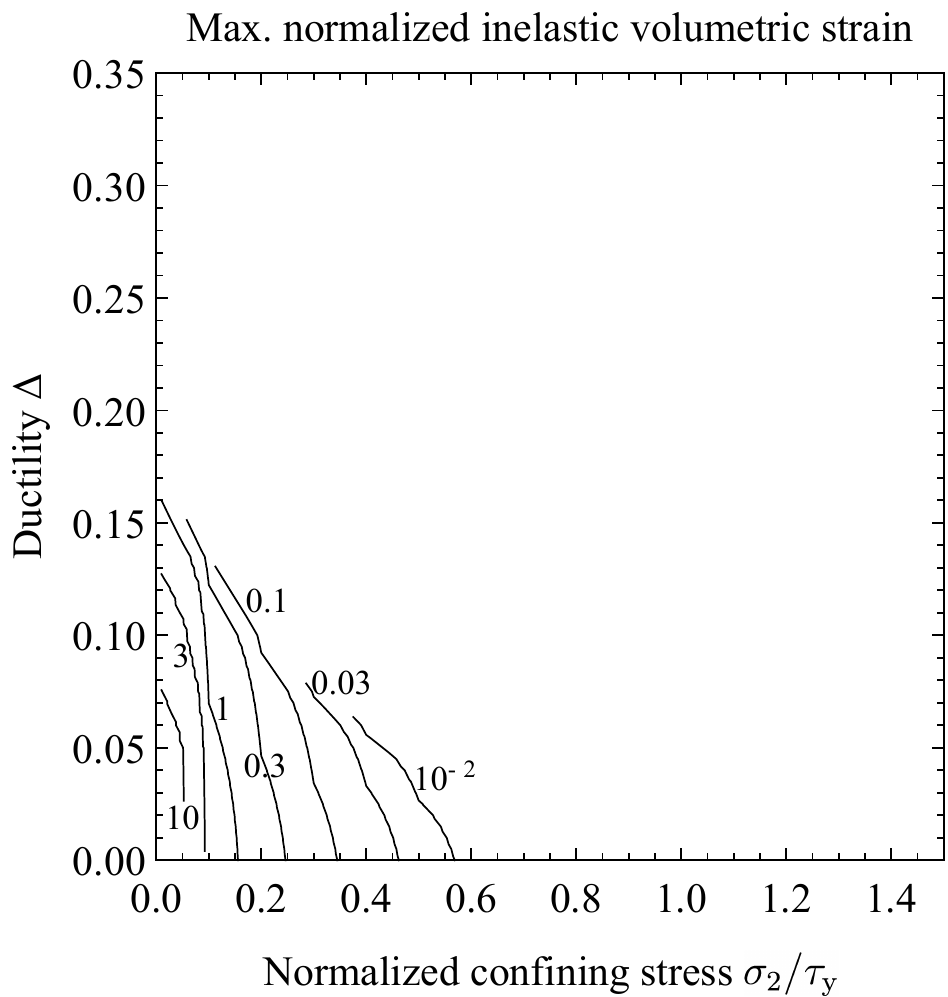}\\
    (a) $\tau_{\mathrm{c}}=0$ & (b) $\tau_{\mathrm{c}}/\tau_{\mathrm{y}}=0.5$
    \end{tabular}
    \caption{Contours of maximum normalized volumetric strains $\epsilon_{v, \mathrm{max}}^{\mathrm{p}} (E^{\prime}/\tau_{\mathrm{y}})/f$ for the case of $\gamma=\theta=\pi/4$, $\mu=0$ with different cohesion.
    %:  (a) $\tau_{\mathrm{c}}/\tau_{\mathrm{y}}=0$ and (b) $\tau_{\mathrm{c}}/\tau_{\mathrm{y}}=0.5$.
    }
    \label{fig:mu0}
\end{figure}

\paragraph{Effect of cohesion}
We show in Fig.~\ref{fig:mu0}  the case of zero friction $\mu=0$ and a non-zero cohesion $\tau_{\mathrm{c}}/\tau_{\mathrm{y}}=0.5$. When the cohesion is zero $\tau_{\mathrm{c}}=0$, a zero friction coefficient should have led to a more brittle behaviour than the case of $\mu=0.1$. However, a much more ductile behaviour is obtained after the introduction of the cohesion $\tau_{\mathrm{c}}=0.5 \tau_{\mathrm{y}}$ (Fig.~\ref{fig:mu0}). 
The cohesion limits the growth of wing cracks and the dilatant volumetric strain by making it more difficult for the preexisting flaw to slide.

\section{Interpretation and analytical estimates}

Our numerical simulations reproduce and extend those provided originally by \cite{HoNe1986}, who only investigated the case of proportional loading and with a single friction coefficient of $\mu=0.4$. The resulting stress-strain behaviour shows systematic strain hardening, where the differential stress asymptotically reaches $2\tau_\mathrm{y}$ at large strain. This upper limit corresponds to the case of infinitely long plastic zones with resolved shear stress $\tau_\mathrm{y}$ oriented at $45^\circ$ from the compression axis. The stability of tensile crack growth in compression inevitably leads to a maximum in tensile crack length (and volumetric strain), beyond which the imposed shear stress leads to unbounded extension of plastic zones.

It is evident from Fig.~\ref{fig:EvContour} and Fig.~\ref{fig:MuEffect} that the knowledge of ductility parameter only is insufficient to determine the limit where dilatancy vanishes. In order to establish a more complete criterion that includes friction coefficient and confining stress, it is useful to establish the stress levels required for significant tensile crack growth vs. the stress where plastic zones dominate the behaviour.

\subsection{Criterion based on cracking onset}

The initiation of tensile cracks is given by the condition \cite[e.g.][Eq. 2.5]{HoNe1986}
\begin{equation} \label{eq:crack_init}
  K_\mathrm{I} = (3/4)\sqrt{\pi c}(\tau_{\hat{x}\hat{y}} - \tau_\mathrm{f})(\sin(\theta/2)+\sin(3\theta/2)) > K_\mathrm{Ic},
\end{equation}
where $\tau_\mathrm{f}$ is the frictional strength on the initial flow. The angle $\theta$ that maximizes $K_\mathrm{I}$ is equal to $0.392\pi$, and the factor $(3/4)(\sin(\theta_\mathrm{c}/2)+\sin(3\theta_\mathrm{c}/2)) \approx \sqrt{3}/2$ to a good approximation. The condition \eqref{eq:crack_init} can thus be rewritten in terms of stress as
\begin{equation} \label{eq:tau_init}
  \tau_{\hat{x}\hat{y}} > \tau_\mathrm{f} + \frac{K_\mathrm{Ic}}{\sqrt{\pi c}}\frac{2}{\sqrt{3}}.
\end{equation}

On the other hand, we know that the shear stress is bounded by $\tau_\mathrm{y}$: if $\tau_{\hat{x}\hat{y}}$ approached $\tau_\mathrm{y}$, plastic zones extend to increasingly large distances ahead of the initial shear crack. Tensile cracks can grow only in the range set by \eqref{eq:tau_init} and $\tau<\tau_\mathrm{y}$. Because tensile crack growth in compression is stable, it requires increasing driving shear stress. Therefore, one may consider that the material is likely more ``plastic'' (i.e, crack growth is not substantial) if the range of shear stress between that at slip initiation on the main crack and the plastic flow stress $\tau_\mathrm{y}$ is narrow. We may define a nondimensional quantity
\begin{equation}
  \Gamma = \left(\tau_\mathrm{f} + \frac{K_\mathrm{Ic}}{\sqrt{\pi c}}\frac{2}{\sqrt{3}}\right)/\tau_\mathrm{y}
\end{equation}
that characterizes the range of possible shear stress in the inelastic regime. When $\Gamma$ is small (near zero), the stress at initiation of tensile cracking is much smaller than the plastic flow stress, so that we expect substantial further crack growth. Conversely, if $\Gamma$ is close to $1$, plastic flow will occur rapidly after the onset of tensile fracturing, and we do not expect much cracking. The frictional strength on the initial flaw $\tau_\mathrm{f}$ depends on the applied load and flaw orientation $\gamma$. A lower bound for $\tau_\mathrm{f}$ is given by the stress level required to initiate frictional slip on the flaw, equal to
\begin{equation} \label{eq:tauf}
  \tau_\mathrm{f}= \left\{
    \begin{array}{ll}
      (\tau_\mathrm{c} + \mu \sigma_2)/(1-\mu) & \text{if }\gamma=\pi/4\\
      (\tau_\mathrm{c} + \mu \sigma_2)/(1-\mu\sqrt(1+\mu^2)+\mu^2) & \text{if }2\gamma=\mathrm{arctan}(1/\mu)
    \end{array}\right.
\end{equation}

We can then rewrite 
\begin{equation}
  \Gamma = \frac{\tau_\mathrm{c} +  \mu \sigma_2}{\tau_\mathrm{y} f(\mu)} + \frac{K_\mathrm{Ic}}{\tau_\mathrm{y}\sqrt{\pi c}}\frac{2}{\sqrt{3}},
\end{equation}
where $f(\mu)$ is either equal to $1-\mu$ (for initial flaws oriented at $45^\circ$ from $\sigma_1$), or to $1-\mu\sqrt(1+\mu^2)+\mu^2$ if we consider optimally oriented flaws. We expect the material to be fully plastic if $\Gamma$ is above a certain critical value. Taking a conservative upper limit of $\Gamma=1$, we obtain a first tentative criterion for fully plastic flow in terms of the ductility parameter, confining stress, friction coefficient and cohesion as follows:
\begin{equation} \label{eq:criterion_0}
  \frac{K_\mathrm{Ic}}{\tau_\mathrm{y}\sqrt{\pi c}} =\Delta>\Delta_\mathrm{crit}  = \frac{\sqrt{3}}{2}\left( 1 - \frac{\tau_\mathrm{c} + \mu \sigma_2}{\tau_\mathrm{y} f(\mu)}\right).
\end{equation}
One obvious limitation of this criterion is that it predicts no dependency on $\sigma_2$ if $\mu=0$. This is not what is observed in fully numerical simulations of the coupled wing crack and plastic zone problem (Fig.~\ref{fig:mu0}).

\subsection{Criterion including some limited tensile cracking}

One way to improve on our criterion is to consider that \emph{some} tensile cracking can occur, but that \emph{dominantly} plastic behaviour (i.e., plastic zone size grow such that $\ell_\mathrm{p}\gg\ell_\mathrm{t}$) occurs if tensile cracks remains small. Here again, we will consider that plasticity is favoured when the range of applied stress between onset of slip and plastic flow stress is narrow. To estimate the stress required for some (small) tensile cracks to develop, we use the approximation of stress intensity factor at the tips of wing cracks in absence of plastic zones \cite[][Eq. 2.6a]{HoNe1986}:
\begin{equation} \label{eq:KI_long}
  K_\mathrm{I} = \frac{2c(\tau_{\hat{x}\hat{y}}-\tau_\mathrm{f})\sin\theta}{\sqrt{\pi(\ell_\mathrm{t}+\ell^*)}} + \sqrt(\pi\ell_\mathrm{t})(\sigma_2 + ((\sigma_1-\sigma_2)/2)(1-\cos2(\theta-\gamma))),
\end{equation}
where $\ell^*=0.27c$ is a regularization length that ensures expressions \eqref{eq:crack_init} and \eqref{eq:KI_long} match in the short wing limit. For simplicity, we assume that $\theta=\gamma$. The growth criterion in terms of shear stress is thus
\begin{equation}
  \tau_{\hat{x}\hat{y}} > \tau_\mathrm{f} + \frac{K_\mathrm{Ic}}{\sqrt{\pi c}}\frac{\pi\sqrt{\ell_\mathrm{t}/c+\ell^*/c}}{2\sin\gamma} + \sigma_2 \frac{\ell_\mathrm{t}}{c}\frac{\pi\sqrt{1+\ell^*/\ell_\mathrm{t}}}{2\sin\gamma}.
\end{equation}
Re-using Equation \eqref{eq:tauf} for the frictional strength, the ductility criterion becomes
\begin{equation}
  \Gamma = \frac{\tau_\mathrm{c}}{\tau_\mathrm{y}f(\mu)} + \frac{\sigma_2}{\tau_\mathrm{y}}\left[ \frac{\mu}{f(\mu)} + \frac{\ell_\mathrm{c}}{c}\frac{\pi\sqrt{1+\ell^*/\ell_\mathrm{t}}}{2\sin\gamma}\right] + \frac{K_\mathrm{Ic}}{\tau_\mathrm{y}\sqrt{\pi c}}\frac{\pi\sqrt{\ell_\mathrm{t}/c+\ell^*/c}}{2\sin\gamma},
\end{equation}
which can be more conveniently expressed as
\begin{equation}
  \Gamma = \frac{\tau_\mathrm{c}}{\tau_\mathrm{y}f(\mu)} + \frac{\sigma_2}{\tau_\mathrm{y}}\left[ \frac{\mu}{f(\mu)} + b(\ell_\mathrm{t}) \right] + a(\ell_\mathrm{t}) \Delta ,
\end{equation}
where $\Delta$ is the ductility index, and $a$ and $b$ are functions of wing length and geometry:
\begin{equation}
  a(\ell_\mathrm{t}) = \frac{\pi\sqrt{\ell_\mathrm{t}/c+\ell^*/c}}{2\sin\gamma}, \quad b(\ell_\mathrm{t}) = \sqrt{\ell_\mathrm{t}/c}\,a(\ell_\mathrm{t}
  ).
\end{equation}
What would be a reasonable choice for $\ell_\mathrm{t}$? This value could be eventually the result of a fit to numerical simulations, but we expect $\ell_\mathrm{t}/c$ to be small for our criterion to make any sense. Setting the value $\ell_\mathrm{t} = \ell^* = 0.27c$ seems appropriate. When $\gamma=\pi/4$, we get $a\approx1.632$ and $b\approx0.848$. Again considering fully plastic flow to be realized when $\Gamma=1$ (as a conservative upper bound), we obtain the following ductility criterion:
\begin{equation} \label{eq:criterion}
  \frac{K_\mathrm{Ic}}{\tau_\mathrm{y}\sqrt{\pi c}} =\Delta>\Delta_\mathrm{crit}  = \frac{1}{a}\left( 1 - \frac{\tau_\mathrm{c}}{\tau_\mathrm{y}f(\mu)} - \left(\frac{\mu}{f(\mu)} +b\right) \frac{\sigma_2}{\tau_\mathrm{y}}\right).
\end{equation}
We see here that introducing a nonzero tensile crack length produces a dependence in pressure even if friction is zero. The prefactor $1/a\approx0.616$ is not significantly different from $\sqrt{3}/2\approx0.866$, but this is by construction: the value of $\ell^*$ is such that the correct short-wing behaviour is predicted by Equation \eqref{eq:KI_long}.

\begin{figure}
\begin{tabular}{cc}
  \centering
  \includegraphics[width=0.45\linewidth]{ 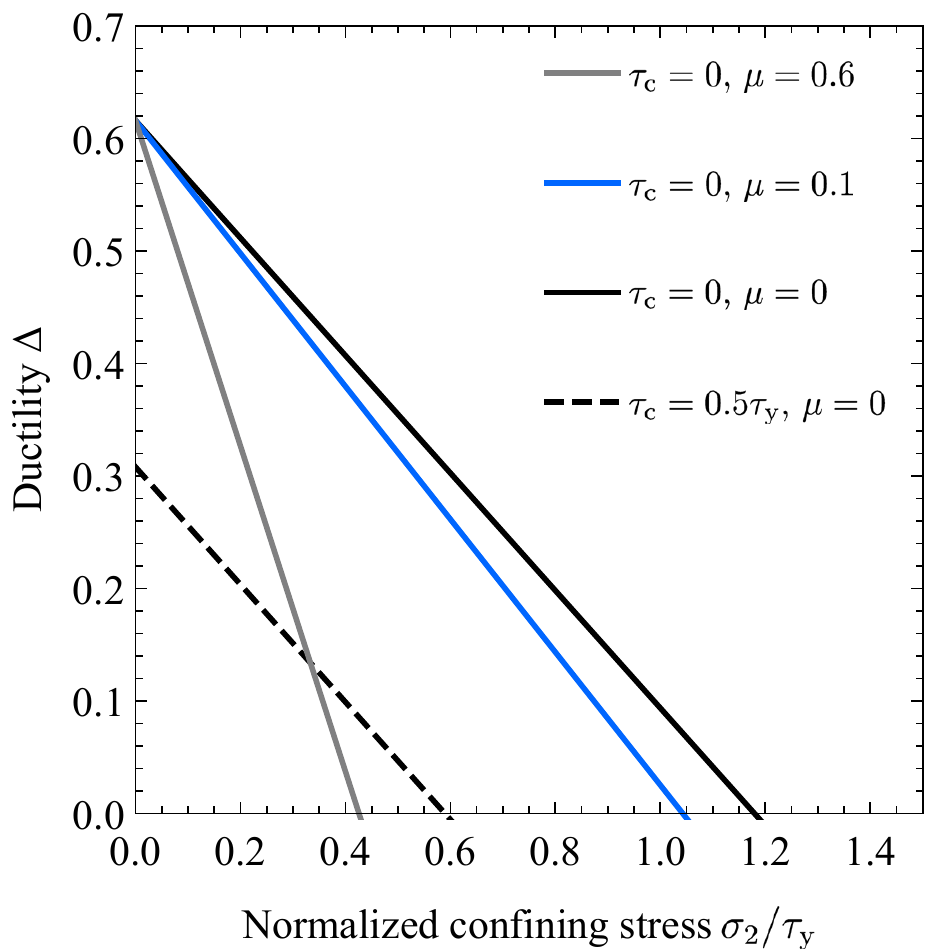} &
  \includegraphics[width=0.45\linewidth]{ 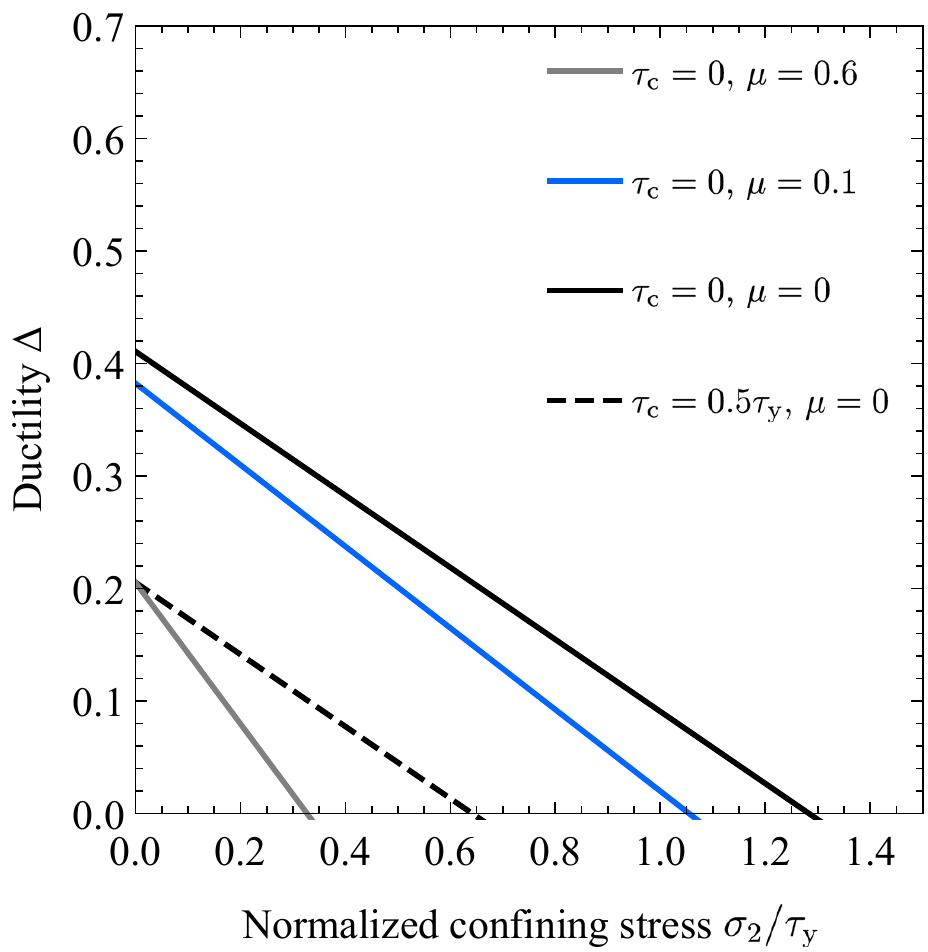}\\
  (a) Criterion \eqref{eq:criterion} & (b) Criterion \eqref{eq:criterion_p}
\end{tabular}
  \caption{Test of (a) the analytical criterion \eqref{eq:criterion} using $a=1.623$ and $b=0.848$ and (b) the analytical criterion \eqref{eq:criterion_p} using  $a=1.623$, $\ell_\mathrm{t}=0.27c$ and $\ell_\mathrm{p}=c$.
  }
  \label{fig:criterion}
\end{figure}

Fig.~\ref{fig:criterion}a illustrates criterion \eqref{eq:criterion} in cases that mimic those tested in the complete numerical model (Fig.~\ref{fig:EvContour} and Fig.~\ref{fig:MuEffect}). Compared to the numerical results, the analytical criterion overpredicts the ductility limit at low pressure (and high ductility). However, at low ductility the prediction is reasonably accurate. Furthermore, the key trends of the numerical simulations are reproduced: lower friction expands the range of pressures where brittle behaviour can occur.

\subsection{Criterion including plastic zones}

We can further improve our analytical criterion (or, at least the match with numerical results) by including considerations on the role of extended plastic zones on the required condition for either the onset of tensile fracturing, or for having only limited growth of tensile cracks. Here we follow again the developments of \cite{HoNe1986} and compute the shear stress applied on the main crack in an initial configuration where there is no tensile cracking and plastic zones have a finite size. The applied differential stress in equilibrium with plastic zones of size $\ell_\mathrm{p}$ is given by \cite[][their Eq. 3.20]{HoNe1986}
\begin{equation}
  \frac{\sigma_1-\sigma_2}{2\tau_\mathrm{y}} = \frac{1 + [(\tau_\mathrm{c} + \mu \sigma_2)/\tau_\mathrm{y} - 1]g(\ell_\mathrm{p}/c)}{\sin(2\gamma) - \mu(1-\cos(2\gamma))g(\ell_\mathrm{p}/c)},
\end{equation}
where $g(x) = (2/\pi)\mathrm{arcsin}\big(1/(1+x)\big)$. Using the above expression in the stress intensity factor \eqref{eq:KI_long} gives
\begin{align}
  \frac{K_\mathrm{I}}{\tau_\mathrm{y}\sqrt{\pi c}} &= \frac{1}{a(\ell_\mathrm{t})(1 -(1-f(\mu))g(\ell_\mathrm{p}/c))}\left[f(\mu)(1-g(\ell_\mathrm{p}/c)) - \frac{\tau_\mathrm{c}}{\tau_\mathrm{y}}(1-g(\ell_\mathrm{p}/c)) \ldots\right. \\
  &\quad \left.- \frac{\sigma_2}{\tau_\mathrm{y}}\left(\mu(1-g(\ell_\mathrm{p}/c)) + \sqrt{\ell_\mathrm{t}/c}(1 -(1-f(\mu))g(\ell_\mathrm{p}/c)) \right) \right].
\end{align}

This stress intensity factor decreases with increasing tensile crack length. By setting a small upper limit for the tensile crack length, for any given plastic zone size we can estimate the maximum stress intensity factor required to maintain tensile cracks smaller than the chosen limit. If toughness is larger than this maximum stress intensity factor, then it will ensure that tensile cracking is indeed limited and the material will effectively remain dominantly governed by the plastic zones.

Obviously, we need to choose somewhat arbitrarily (i) the upper bound for tensile crack length, and (ii) the plastic zone size at which we want to test the material's ``plasticity''. As before, we can suspect that we need a small but nonzero $\ell_\mathrm{t}$, and here again we shall settle on $\ell_\mathrm{t}=\ell^*$. How to choose $\ell_\mathrm{p}$? One way to imagine a reasonable option is to consider a case where imposed stress has led to significant growth of the plastic zones in absence of cracking (i.e., the material has yielded in a ``plastic'' way), and ask ourselves whether tensile cracking will occur at any further point of the deformation. Thus, we should take a large enough value of $\ell_\mathrm{p}$ so that the material has already experienced significant plasticity. A first guess is to use $\ell_\mathrm{p}/c=1$, which represents a situation where plastic zones are commensurate to initial flaw size. Eventually, our choices of tensile crack and plastic zone sizes need to evaluated by comparison with numerical results.

Using $\ell_\mathrm{p}=c$ gives the following simplified expression for our ductility criterion:
\begin{equation} \label{eq:criterion_p}
  \frac{K_\mathrm{Ic}}{\tau_\mathrm{y}\sqrt{\pi c}} =\Delta>\Delta_\mathrm{crit}  = \frac{1}{a(1+f(\mu)/2)}\left( f(\mu) - \frac{\tau_\mathrm{c}}{\tau_\mathrm{y}} - \left(\mu + \big(1+f(\mu)/2\big)\sqrt{\ell_\mathrm{t}/c}\right) \frac{\sigma_2}{\tau_\mathrm{y}}\right).
\end{equation}
The criterion is plotted in Fig.~\ref{fig:criterion}b. It is seen to be an excellent predictor of the numerical simulations, especially at low confining pressure. Increasing $\ell_\mathrm{p}$ improves somewhat the match at low ductility, but worsens it at low pressure.

\subsection{Confining pressure threshold for fully plastic behaviour}

The three criteria \eqref{eq:criterion_0}, \eqref{eq:criterion} and \eqref{eq:criterion_p} provide us with relationships between ductility, friction, cohesion and confining stress necessary for the dominance of plasticity and limited tensile fracturing. Goetze's criterion states that fully plastic, nondilatant deformation is achieved when $(\sigma_1-\sigma_2)<\sigma_2$. In the context of our model, the upper bound for differential stress is $2\tau_\mathrm{y}$ (for flaws oriented at $45^\circ$ from compression axis). With this upper bound, Goetze's criterion is equivalent to $2<\sigma_2/\tau_\mathrm{y}$. 

Most rocks at low temperature have a high strength and low toughness, so that their ductility index is close to zero. In the limit $\Delta=0$, the criteria determined above can be rewritten in terms of minimal confining stress $\sigma_2/\tau_\mathrm{y}$ required for plastic behaviour:
\begin{equation}
  \sigma_2/\tau_\mathrm{y} > \left\{
    \begin{array}{l}
      \left((1-2\Delta/\sqrt{3})f(\mu) - \tau_\mathrm{c}/\tau_\mathrm{y} \right)/\mu, \quad \text{crit. \eqref{eq:criterion_0}}\\
      \left((1-a\Delta)f(\mu) - \tau_\mathrm{c}/\tau_\mathrm{y} \right)/(\mu+bf(\mu)), \quad  \text{crit. \eqref{eq:criterion}}\\
      \left(f(\mu)-a(1+f(\mu)/2)\Delta - \tau_\mathrm{c}/\tau_\mathrm{y} \right)/(\mu + (1+f(\mu)/2)\sqrt{\ell_\mathrm{t}/c}) \quad \text{crit. \eqref{eq:criterion_p}}
    \end{array}
  \right.  \nonumber
\end{equation}

\begin{figure}
  \centering
  \includegraphics[width=0.45\linewidth]{ 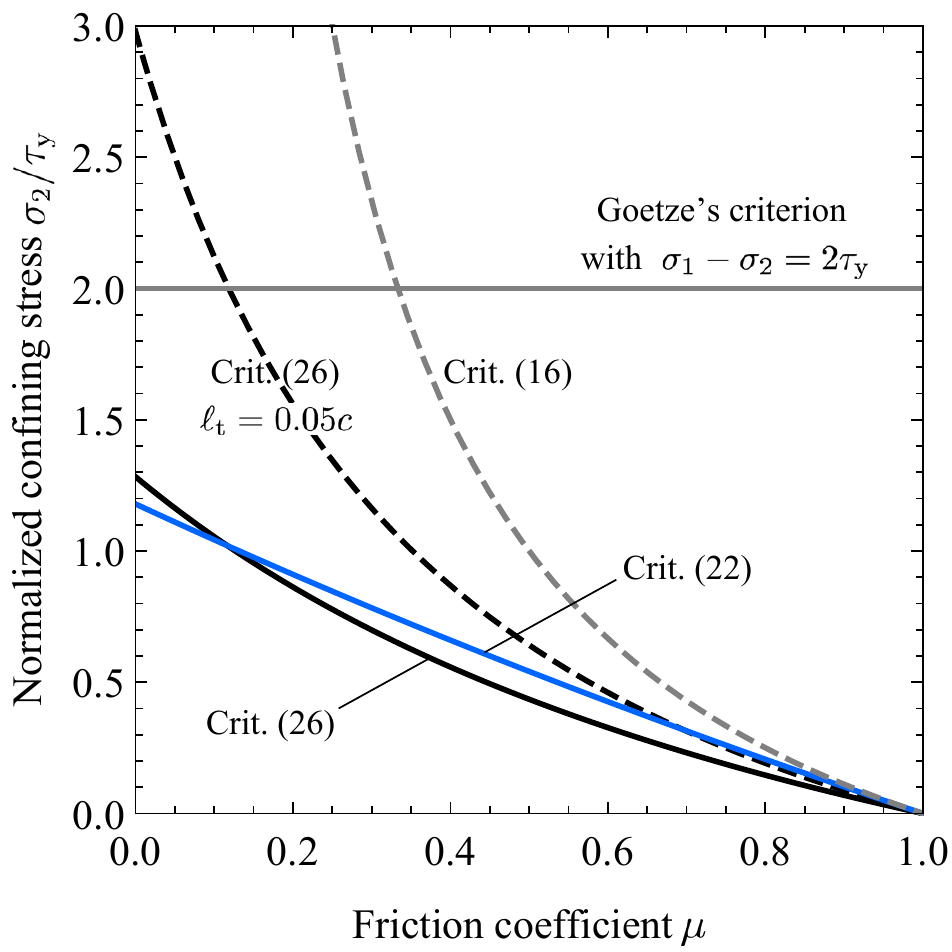}
  \caption{Limit in confining stress above which fully plastic behaviour is expected, using our three possible criteria. Computation done using $\Delta=0$, $\tau_\mathrm{c}=0$, $\ell_\mathrm{t}=\ell^*=0.27c$ and $\ell_\mathrm{p}=c$, except otherwise noted.}
  \label{fig:pc_limit}
\end{figure}

These criteria are shown in Fig.~\ref{fig:pc_limit}. For the most extreme case of $\mu=0$, the criteria derived from \eqref{eq:criterion} and \eqref{eq:criterion_p} both reach a maximum $\sigma_2/\tau_\mathrm{y}\approx 1.2$, which is below Goetze's criterion. In other words, Goetze's criterion is stricter than what we derived in our model, which is consistent with the assumption that there is modest but nonzero tensile fracturing.  Reducing the size of tensile cracks significantly increases the confining stress limit, and both criteria become closer to the end-member case given by \eqref{eq:criterion_0}, which is equivalent to the classic case where stress is limited by friction only; here, Goetze's criterion is equivalent to setting $\mu\approx 0.3$ \cite{FrEv1990}. For larger ductility, this friction threshold is slightly smaller.

\section{Discussion}

\subsection{Comparison between numerical results and experimental data}

In this section, we aim to compare the rock strength obtained in numerical simulations at different confining stresses with laboratory data. As discussed above, the model always predicts an ultimate strength given by $2\tau_\mathrm{y}$ at large strain (Fig.~\ref{fig:Illustration}b), which is due at least in part to the absence of interactions between cracks. Nevertheless, we can use the model predictions at low strain (i.e., for small plastic zones and/or tensile crack length) to determine trends in the resulting strength.

For this purpose, we estimate the rock strength as the differential stress $(\sigma_1-\sigma_2)$ achieved at a given total strain of $\epsilon_t=\epsilon_{\sigma}$, where $\epsilon_{\sigma}$  is a constant, small strain limit that we pick to match experimental data. The rock strength obtained this way depends on the confining stress $\sigma_2$, shear yielding stress $\tau_{\mathrm{y}}$, elastic modulus $E^{\prime}$,  fracture toughness $K_{\mathrm{Ic}}$, flaw size $2c$, and crack density $f$, where $E^{\prime}$ and $\tau_{\mathrm{y}}$ are temperature-dependent properties and should ideally correspond to single crystal properties.

\begin{figure}
    \centering
    \begin{tabular}{cc}
    \includegraphics[height=0.45 \linewidth]{ 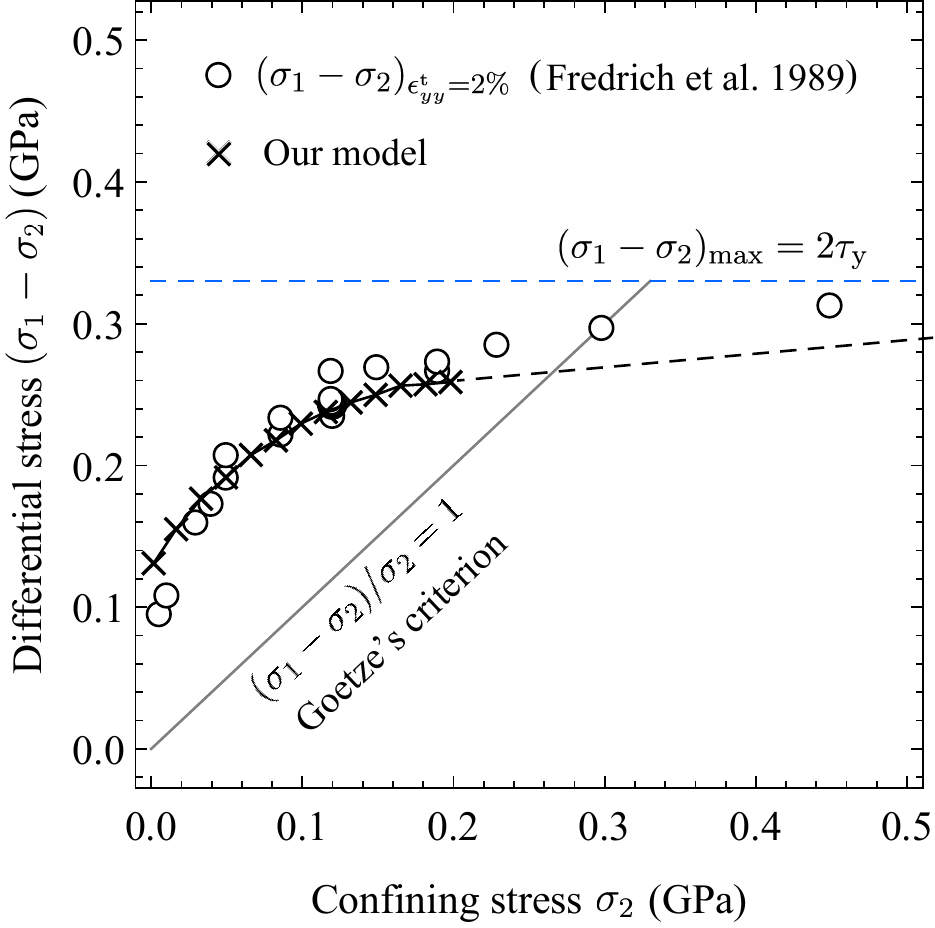} & 
    \includegraphics[height=0.45 \linewidth]{ 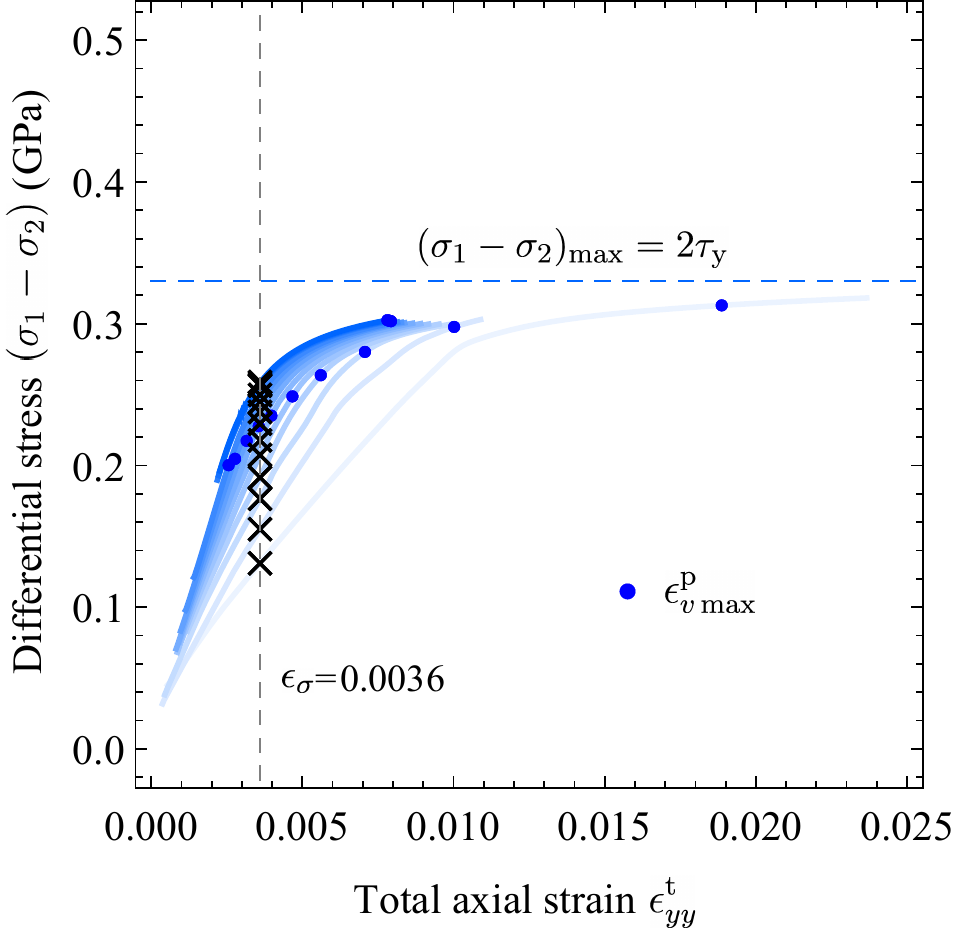}\\
    (a) & (b)
    \end{tabular}
    \caption{Plots of (a) differential stress versus confining stress for Carrara marble at room temperature \cite{FrEv1989}, and (b) differential stress versus the strain obtained from the numerical modelling. The numerical results are obtained by setting $\gamma=\theta=\pi/4$, $\mu=0.1$, and $\tau_{\mathrm{c}}/\tau_{\mathrm{y}}=0$.}
    \label{fig:Marble}
\end{figure}

We compare the model output with laboratory triaxial data obtained in Carrara marble \cite{FrEv1989} at room temperature, which will serve as an illustrative example. We follow \cite{FrEv1990} and use $\tau_\mathrm{y}=165$~MPa, $K_\mathrm{Ic}=0.195$~MPa~m$^{1/2}$, and $E=84$~GPa, $\nu=0.26$ \cite{NiFo2017}. Using a friction coefficient of $\mu=0.1$, a crack density $f=0.5$ and a strain $\epsilon_\sigma=0.36$\%  allows us to reproduce the main trend of the laboratory data (Fig.~\ref{fig:Marble}): at a given strain, increasing confining pressure leads to an increase in differential stress, but this increase saturates at high pressure.

% horizontal line
% two other lines related to Goetze's criterion

\begin{figure}
    \centering
    \begin{tabular}{cc}
    \includegraphics[height=0.45 \linewidth]{ 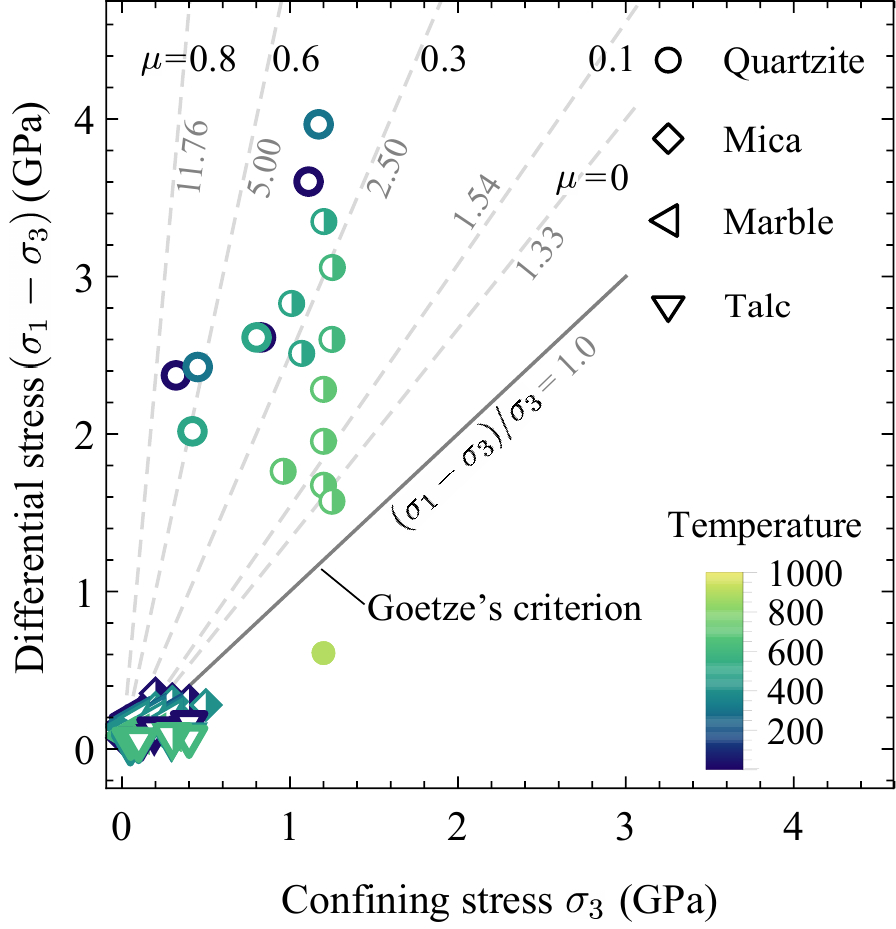} & 
    \includegraphics[height=0.45 \linewidth]{ 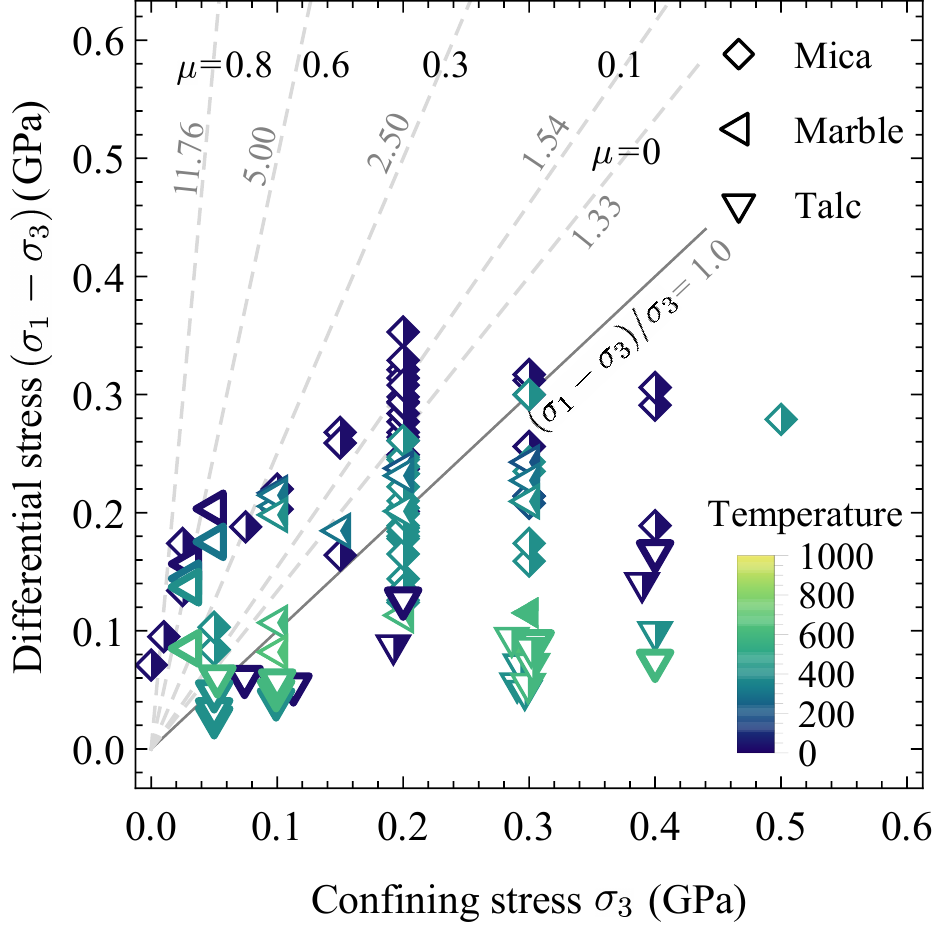}\\
    (a) & (b)
    \end{tabular}
    \caption{Plots of material strength (differential stress) versus confining stress for quartzite \cite{HiTu1994}, marble \cite{Ryba2021,FrEv1989}, talc \cite{Esca2008} and mica schist \cite{ShKr1992} at different temperatures with a strain rate of $\dot{\epsilon}\sim 10^{-5}$. The ultimate strength of the specimen with brittle failure is marked with empty symbols. The stresses corresponding to the knee of the stress-strain curve are plotted with solid and semi-solid symbols for the plastic and transitional regime, respectively. The panel (b) is the subset of the panel (a) zooming into a lower range of the confining stress. The dashed lines correspond to the confining stress ratios $(\sigma_1-\sigma_3)/\sigma_3$ at which the normalized inelastic volumetric strain $\epsilon_{v, \mathrm{max}}^{\mathrm{p}} (E^{\prime}/\tau_{\mathrm{y}})/f$ obtained from the sliding crack model goes below 0.2\% for $\Delta=0$ with different friction coefficients $\mu=0, \, 0.1, \, 0.3, \, 0.6, \, 0.8$. 
    }
    \label{fig:GoetzeData}
\end{figure}

Despite this qualitative agreement, the stress-strain curves predicted by the model are difficult to reconcile quantitatively with experimental data if we consider $\tau_\mathrm{y}$ to be representative of single crystal properties. This is somewhat expected since the bulk strength of polycrystals in the semi-brittle regime results from the combination of possibly several slip systems, in addition to grain boundary sliding and microcrack interactions which are not accounted for in the model. 

The model is however useful to test how physical parameters influence the transition from dilatant to nondilatant deformation. In their comprehensive study on Carrara marble, \cite{Ryba2021} use the original upper bound for ductility $\Delta\approx0.26$ reported by \cite{HoNe1986}, and conclude that at temperature above $200^\circ$C, ductility ranges from $0.28$ to $1.74$ and the wing crack model predicts that Carrara marble is always fully plastic, which is at odds with their experimental data. The  occurrence of microcracking, or at least of some pressure-dependency of strength, up to elevated temperature ($800^\circ$C at strain rates above $10^{-4}$~s$^{-1}$) and modest pressure ($<300$~MPa) can be reconciled with the model if we assume that friction coefficient is much smaller than the value of $0.4$ used by \cite{HoNe1986}. Indeed, the limit for $\Delta$ to achieve fully plastic flow regardless of confining stress increases with decreasing friction, up to a maximum of about $0.35$. If we further consider that fully plastic flow strictly means that the tensile cracks do not even initiate, then we can use the criterion \eqref{eq:criterion_0} and look for a combination of ductility and friction coefficient that will match the laboratory observations. In this case, the maximum ductility above which fully plastic behaviour is reached, regardless of friction coefficient, is $\Delta=\sqrt{3}/2 \approx 0.87$. In other words, the presence of tensile microfractures in marble at elevated temperature might indicate that friction between preexisting defects, such as grain boundaries, is low. Although the model geometry is too simplified to really represent any actual microstructure, the hypothesis of low friction at gain boundaries is consistent with the microstrain measurements of \cite{QuEv2016}, which show direct evidence of localized shear strains across grain boundaries in Carrara marble samples deformed at $400^\circ$C.

In quartzite, \cite{HiTu1994} report that Goetze's criterion is appropriate to describe the transition from semi-brittle to fully plastic flow. However, in the semi-brittle regime at temperatures above $500^\circ$C and pressures above $1$~GPa, \cite{HiTu1994} also note that deformation is dominated by shear cracks rather than by tensile cracks, even though Goetze's criterion is not necessarily satisfied. This behaviour appears at $\sigma_1 - \sigma_2 \leq 2\sigma_2$, which is consistent with our simple criterion \eqref{eq:criterion} (in the limit of $\Delta=0$, assuming $\sigma_1-\sigma_2=2\tau_\mathrm{y}$) if friction is small, $\mu\approx0.2$ (Fig.~\ref{fig:pc_limit}). Assuming the more strict criterion \eqref{eq:criterion_0}, the dominance of shear over tensile cracking is consistent with a friction coefficient of $0.6$ (Fig.~\ref{fig:pc_limit}), which coincides with Byerlee's rule \cite{Byer1978}. Increasing ductility would restrict the range of possible friction coefficients compatible with the observed behaviour. 

Unfortunately, constraining the ductility parameter is challenging for quartz: while fracture toughness is relatively well known \cite[e.g.][]{Atki1984,DaGu1985}, the yield stress in shear is not well constrained. At any given temperature, there are considerable variations in critical resolved shear stresses for slip systems in quartz as functions of water activity and pressure \cite[e.g.][]{BlCh1984,DoTr1985}. In synthetic wet quartz at $583$ to $590^\circ$C, \cite{DoTr1985} report yield stresses ranging from 100 to 200~MPa. Using a fracture toughness of $K_\mathrm{Ic}\approx 0.2$~MPa~m$^{1/2}$ and a flaw size of $100$~$\mu$m (on the lower end of the grain size of the rock tested by \cite{HiTu1994}) leads to a possible upper bound of $\Delta\approx0.06$ to $0.11$. By contrast, the yield stress in natural dry quartz at $600^\circ$C is of the order of $4$~GPa \cite{BlCh1984}, and taking this value makes ductility effectively zero. 

The role of low friction in promoting tensile cracking is best illustrated by the case of phyllosilicates. In talc aggregates, \cite{EdPa1972} report dilatancy at pressures largely exceeding the shear strength at room temperature: the pressure required to achieve nondilatant deformation is around $600$~MPa, for a strength of around $200$~MPa. At elevated temperatures up to $700^\circ$C, where talc becomes chemically unstable, \cite{Esca2008} also report pervasive microcracking regardless of the confining pressure (up to $300$~MPa). The pressure dependence of strength in talc shows that friction coefficient is of the order of $0.1$. The fracture toughness and yield stress of talc are extremely anisotropic, and precise data are missing to convincingly estimate the ductility parameter. In the limit of $\Delta=0$, which could correspond to a vanishingly small toughness along the cleavage plane of talc single crystals, the pressure required to achieve negligible dilation is of the order of $1.2\tau_\mathrm{y}$ (Fig.~\ref{fig:MuEffect}). Assuming the strict criterion \eqref{eq:criterion_0} where tensile cracks do not nucleate, a friction coefficient of $0.1$ and negligible ductility imply a pressure threshold of $9\tau_\mathrm{y}$. Using an end-member yield stress of the order of $100$~MPa at room temperature (half the macroscopic differential stress), the pressure threshold is of the order $900$~MPa, which is broadly consistent with \cite{EdPa1972}'s results.

In mica schist, \cite{ShKr1992} also report the presence of microcracks at elevated temperature and pressures up to $500$~MPa, even when the strength was below the imposed confining pressure. The persistence of dilatancy at high pressure in mica schist was attributed at least in part to the propensity of mica crystals to form kink bands, which produce opening at grain boundaries and intracrystalline delamination. Similar observations were made in talc by \cite{Esca2008}. In both mica and talc, kink band formation is facilitated by the strong plastic anisotropy of the minerals, which have one easy glide system along the basal plane. Although the wing crack model is not designed to reproduce the kinking mechanism, it is based on similar physical assumptions, with a prominent role for slip (captured by the combination of friction and cohesion) and a coupling between slip and tensile cracking. Detailed models of crack growth induced by dislocation pile ups and dislocation wall splitting have indeed similar mathematical forms to that of the wing crack model \cite{FrWi1968,Wong1990}. Thus, the qualitative relationship between low friction (or easy glide in one plane) and the persistence of microcracking up to high pressure is likely a generic feature.

\subsection{Other effects and possible model improvements}

\paragraph{Effect of pore pressure}
The presence of pore pressure will change the effective normal stress but not the shear stress. Assuming that the fluid is non-viscous, another dimensionless parameter $p_\mathrm{f}/\tau_{\mathrm{y}}$ gets involved in the brittle-plastic transition process. Under drained conditions, the pore pressure is uniform throughout the system. The governing equations (Eq.~\ref{eq:governingwing} and Eq.~\ref{eq:governingplastic}) and boundary conditions (Eq.~\ref{eq:boundarywing} and Eq.~\ref{eq:boundaryplastic}) will remain valid when we replace the total/bulk stresses with the effective ones $\sigma^{\prime}=\sigma-p_\mathrm{f}$, as long as frictional strength is also governed by Terzaghi's effective stress. As a result, we obtain the same contours of volumetric strains as shown in Fig.~\ref{fig:EvContour} except that the x-axis turns to the effective confining stress $(\sigma_2/\tau_{\mathrm{y}}-p_\mathrm{f}/\tau_{\mathrm{y}})$. Under the same confining stress $\sigma_2/\tau_{\mathrm{y}}$, the material turns to be more brittle due to the presence of pore pressure. 

In the model, the pore pressure plays a first order role in the strength at small strain, where plastic zones are limited in size: pore pressure decreases the frictional strength on the initial flaw, and acts against the confining stress on the tensile cracks. In this regime, the effective stress law for strength is valid, and the effective stress coefficient is equal to 1. When plastic zones grow significantly, the overall strength is mostly dictated by the shear stress $\tau_\mathrm{y}$, and fluid pressure has a diminishing impact. In this regime, strength becomes gradually independent of both confining and pore pressure, but it is still the simple difference $\sigma_2-p_\mathrm{f}$ that exerts a control on the behaviour. In other words, the effective stress coefficient remains equal to 1.

If the effective stress law for frictional strength on the initial flaw changes, as suggested by \cite{HiBe2015,Beel2016} on the basis of adhesion theory, we expect the onset of microfracturing to be impacted. However, the impact of fluid pressure on tensile crack growth still involves the direct difference between confining and fluid pressure, which is the net normal stress applied on the open microcracks. Thus, even in the case of pressure-independent slip on the initial flaw (e.g., accounting only for the cohesive term $\tau_\mathrm{c}$), the difference $\sigma_2-p_\mathrm{f}$ should remain a key control parameter as soon as tensile cracks start growing.

\paragraph{Rate dependency}
The model presented here is time independent: we followed the original assumptions of \cite{HoNe1986} and considered only the stress dependency of friction, crack growth and plastic zone extension. In practice, all these processes are expected to be rate-dependent.

Friction coefficient is likely weakly rate-dependent if we include the possibility that rate-and-state friction laws apply at the small scale considered in the model. Tensile fracture growth is also time-dependent, and slow crack growth occurs at $K_\mathrm{I}$ less than $K_\mathrm{Ic}$ due to stress corrosion mechanisms \cite[e.g.][]{Atki1984}. A slower wing crack growth would lead to a smaller apparent toughness and a more brittle behaviour. This effect was analysed in detail in the brittle regime (in absence of plastic zones) by \cite{BrBa2012,Mall2015}. In the presence of significant plastic deformation, however, we expect the dominant time-dependent effect to be due to the rate-sensitivity of the yield stress $\tau_\mathrm{y}$ \cite{EvKo1995}. Indeed, elevated strain rate tends to promote more brittle, dilatant behaviour, as observed systematically in calcite \cite{Rutt1974,Ryba2021}. Using a power law creep relationship for the yield stress $\dot{\epsilon}=B\tau_\mathrm{y}^n$, it is straightforward to rewrite all our criteria in terms of a critical normalized strain rate $\dot{\epsilon}/\dot{\epsilon}_\mathrm{c}$ above which significant microcrack dilatancy is expected, where 
\begin{equation}
  \dot{\epsilon}_\mathrm{c} = B\left(\frac{K_\mathrm{Ic}}{\sqrt{\pi c}}\right)^n
\end{equation}
is a characteristic strain rate. This operation is similar in spirit to the brittle-ductile transition model of \cite{ReSc2001}. Although their approach was more focused on predicting peak stress and failure, they also derive a critical strain rate in the form $\dot{\epsilon}/\dot{\epsilon}_\mathrm{c}$ (with the exact same expression for $\dot{\epsilon}_\mathrm{c}$) based on plastic relaxation of elastic stress concentration at crack tips, with a threshold that depends on the applied stress and friction coefficient.

While quantitative predictions are difficult to achieve, and flow law parameters are also highly uncertain especially at low temperature, the characteristic strain rate $\dot{\epsilon}_\mathrm{c}$ is likely a useful quantity that is not strongly model-dependent and can be used to scale experimental data and help their interpretation, as shown in detail by \cite{ReSc2001}.

\paragraph{Size-dependent effects}
The model contains only one length scale that normalizes all physical dimensions: the initial flaw size $2c$. In wing crack models, this size is commonly chosen to be commensurate or equal to the grain size \cite[e.g.][]{AsSa1990}, so that we envision the initial flaw to be a grain boundary or a similar grain-scale feature. The ductility index $\Delta$ scales with $1/\sqrt{c}$, which should make materials of larger grain size more brittle. However, this is often not the case: \cite{FrEv1990} report that calcite rocks tend to be more brittle with \emph{decreasing} grain size, and they suggest that the wing crack model can be reconciled with laboratory observations by considering how size effects impact other quantities such as fracture toughness and plastic yield stress. In particular, the shear yielding stress $\tau_{\mathrm{y}}$ is known to be grain size-dependent. For instance, recent nano-indentation experiments have revealed that the strength of single crystal and polycrystalline olivine decreases with increasing grain size \cite{Kuma2017}. As noted by \cite{FrEv1990}, a Hall-Petch relationship of the type $\tau_\mathrm{y}\propto\sqrt{1/c}$ would only predict size-independent ductility parameter, and one would require either a stronger dependency of $\tau_\mathrm{y}$ on size or that toughness also increases as $\sqrt{c}$ to properly match calcite data. 

However, the sole focus on the ductility index $\Delta$ provides limited insights. Here we highlight the existence of a \emph{critical} $\Delta_\mathrm{crit}$ above which tensile cracking becomes negligible. The threshold $\Delta_\mathrm{crit}$ itself depends on friction coefficient, confining stress, and on $\tau_\mathrm{y}$ (see approximate criteria \ref{eq:criterion_0}, \ref{eq:criterion} and \ref{eq:criterion_p}). In all cases, using a Hall-Petch relation where $\tau_\mathrm{y}\propto1/\sqrt{c}$ leads indeed to a size-independent ductility index, but also to an increase in $\Delta_\mathrm{crit}$ with decreasing grain size, which is consistent with a trend towards a more brittle behaviour in fine-grained materials.

\paragraph{Interactions between cracks and plastic zones} 
One key assumption of the model is that we considered each wing crack and extended plastic zones to be independent, i.e., interactions between neighbouring cracks are neglected. Isolated tensile cracks grow stably in compressive stress fields: an ever increasing differential stress is required to make them grow. In this sense, microcrack growth is a strain-hardening mechanism. By contrast, plastic zones grow unbounded when the applied shear stress reaches $\tau_\mathrm{y}$. Thus, the stress-strain curve is dominated by the growth of plastic zones at large strain, regardless of the length of tensile cracks. However, this situation is not realistic: if tensile cracks grow to sizes commensurate to the spacing between wing crack elements, they will interact and likely induce a decrease in hardening or even a weakening \cite[e.g.][]{AsSa1990} at differential stresses below $2\tau_\mathrm{y}$. It is likely that crack interactions are one of the key missing effect that limits the applicability of the model and the validity of stress-strain curves at large strain; however, accounting for interactions is a significant challenge that would require elaborate numerical solutions and involve more geometrical assumptions (e.g., relative position of cracks, detailed crack path computations). The utility of the wing crack model resides in the simplicity of its assumptions (regardless of the complexity of the mathematical treatment), which allowed us to establish simple bounds for the onset of nondilatant deformation.

\section{Conclusions}
We investigated the brittle-plastic transition by revisiting the frictional sliding crack model due to \cite{HoNe1986}. Assuming a constant confining pressure, similarly to typical loading conditions in the laboratory, we account for both the development of tensile wing cracks and plastic deformation across the brittle-plastic transition, and estimate the stress-strain evolution. Numerical simulations and analytical estimates indicate that the onset of nondilatant deformation results from the interplay between the ductility index $\Delta=K_\mathrm{Ic}/\sqrt{\pi c}$, originally defined by \cite{HoNe1986}, and a combination of other physical quantities (e.g., Equation~\ref{eq:criterion_p}). These include the confining stress, shear yielding stress, defect size, friction coefficient and cohesion. In particular, low friction leads to a significantly extended brittle regime, which may explain the brittle failure in minerals like talc at high confining stress. 

Theoretical predictions are in qualitative agreement with many experimental observations regarding the relation between material strength and confining pressure. The model however underpredicts the pressure required to suppress dilatancy unless very low friction coefficients are used. This misestimation possibly results from the limits of the strength predictions due to the systematic $2\tau_\mathrm{y}$ asymptote at large strain. 
Goetze's criterion, although remaining a rule of thumb, seems to better approximate the confining stress at the brittle-ductile transition. However, the effectiveness of Goetze's criterion probably results from the low ductility index and medium friction coefficient shared by most rock forming minerals \cite{Byer1978}, and needs to be dealt with caution for rocks containing sliding interfaces with low friction.

\section*{Acknowlegements}
Funding by the European Research Council under the European Union's Horizon 2020 research and innovation programme (project RockDEaF, grant agreement \#804685 to N.B.) and the UK Natural Environment Research Council (grants NE/K009656/1, NE/M016471/1 and NE/S000852/1) is acknowledged.

\section*{Data availability\label{sec:data}}
The numerical code used in this paper is available on the Zenodo platform with the identifier\\~\url{https://doi.org/10.5281/zenodo.7304196}.

% \section*{Credit authors contributions}
% Dong Liu: Methodology, Software, Validation, Formal Analysis, Visualization, Writing-Original Draft.
% Nicolas Brantut: Conceptualization, Methodology, Formal Analysis, Supervision, Resources, Writing-Review \& Editing 

%\printbibliography
%\bibliographystyle{plain}

\appendix

\section{Exact formulation \label{ax:Formulation}} 

Following \cite{HoNe1986}, we establish the governing equations in the complex plane of $\hat{x}+\mathrm{i}\hat{y}$, and compression is viewed negative for stresses in all appendix. We denote $z$, $z_o$ to represent respectively the collocation nodes and the nodes of unknowns along the wing cracks and plastic zones.
\begin{equation}
\begin{aligned}
    &z=c + s\, \mathrm{e}^{\mathrm{i} \theta}, \quad z_o=c + r\, \mathrm{e}^{\mathrm{i} \theta}, \quad s, r \in(0, \ell_{\mathrm{t}}), \quad \text{for wing cracks}\\
    &z=c + u, \quad z_o=c+t, \quad u,  t \in (0, \ell_{\mathrm{p}})  \quad \text{for plastic zones}
\end{aligned}
\end{equation}
where $\mathrm{i}=(-1)^{1/2}$, and $\mathrm{e}^{\mathrm{i} \theta}=\cos\theta+\mathrm{i} \sin \theta$.

The dislocation density can be expressed as the sum of a real part and an imaginary part, which indicates respectively the component along the $\hat{y}$ direction (normal to the sliding crack) and that along the $\hat{x}$ direction (parallel to the sliding crack). 
We define $\alpha_{\mathrm{t}}$ as the dislocation density on the tensile wing crack $PQ$, and $\alpha_{\mathrm{p}}$ the dislocation density along the plastic zone $PR$.

\begin{equation}
\begin{aligned}
    \alpha_{\mathrm{t}}=\frac{E^{\prime} \mathrm{e}^{\mathrm{i} \theta}}{8 \pi \mathrm{i}} (\frac{\partial [v_{\hat{r}}]}{\partial r}+\mathrm{i}\frac{\partial[v_{\hat{\theta}}]}{\partial r})\\
    \alpha_{\mathrm{p}}=\frac{E^{\prime}}{8 \pi} \left(-\mathrm{i}\frac{\partial [v_{\hat{x}}]}{\partial t}\right)
\end{aligned}
\end{equation}
where $E^{\prime}=E/(1-\nu^2)$ is the plane strain elastic modulus, $\nu$ Poisson's ratio, and  $[v]$ the dislocation on the crack surface.

\paragraph{For wing cracks}
No displacement is allowed normal to the slip line at the kink point $P$, $P^{\prime}$. Eq.~(\ref{eq:boundarywing}) thus turns to
\begin{equation}
    \int_0^{\ell_{\mathrm{t}}} (\alpha_{\mathrm{t}} +\overline{\alpha_{\mathrm{t}}}) \text{d}r=0
    \label{eq:Geqns1}
\end{equation}
The crack surfaces $PR$ and $P^{\prime}R^{\prime}$ fulfills the stress-free conditions, and the singular integral form of Eq.~(\ref{eq:governingwing}) becomes
\begin{equation}
\begin{aligned}
       2\int_0^{\ell_{\mathrm{t}}} \frac{\overline{\alpha_{\mathrm{t}}} \mathrm{e}^{\mathrm{i} \theta}}{s-r}\text{d}r+\int_0^{\ell_{\mathrm{t}}}L(c+s\, \mathrm{e}^{\mathrm{i} \theta}, c+r\, \mathrm{e}^{\mathrm{i} \theta}; \alpha_{\mathrm{t}}, \theta)\text{d}r+\\
       \int_0^{\ell_{\mathrm{p}}} \alpha_{\mathrm{p}} K_1(s, t)\text{d}t+\int_0^{\ell_{\mathrm{p}}} \overline{\alpha_{\mathrm{p}}} K_2(s, t)\text{d}t+
       \int_0^{\ell_{\mathrm{p}}} L(c+s\, \mathrm{e}^{\mathrm{i} \theta}, c+t; \alpha_{\mathrm{p}},\theta) \text{d}t+\\
       (\tau_{\hat{x}\hat{y}}^{\infty}-\mu \sigma_{\hat{y}}^{\infty}+\tau_{\mathrm{c}}) S(c+s\, \mathrm{e}^{\mathrm{i} \theta}, \theta) +(\sigma_{\hat{\theta}}^{\infty}+\mathrm{i} \tau_{\hat{r} \hat{\theta}}^{\infty})_{\mathrm{t}}=0
\end{aligned}
\label{eq:Geqns2}
\end{equation}
with
\begin{equation}
    K_1(s, t)=\frac{1}{s\, \mathrm{e}^{\mathrm{i} \theta} -t}-\frac{(s\, \mathrm{e}^{-\mathrm{i} \theta}-t)e^{2 i \theta}}{(s\, \mathrm{e}^{\mathrm{i} \theta}-t)^2}, \quad K_2(s, t)=\frac{1}{s\, \mathrm{e}^{-\mathrm{i} \theta} -t}+\frac{e^{2 i \theta}}{s\, \mathrm{e}^{\mathrm{i} \theta}-t}
\end{equation}

$L(z,z_o; \alpha,\theta)$ represents the stress generated at $z$ due to the distributed dislocation density of $\alpha$ at $z_o$. It can be written in two parts associated respectively with the dislocation density $\alpha$ and its conjugate. 

\begin{equation}
    L(z, z_o; \alpha, \theta)=\alpha L_1(z, z_o; \theta)+\overline{\alpha} L_2(z, z_o; \theta)
\end{equation}
\begin{equation}
    \begin{aligned}
    & L_1(z, z_o; \theta)=\beta U\{J(z, z_o)+J(z, \overline{z_o}); \theta\}+\overline{\beta} U\{(z_o-\overline{z_o})G(z, z_o); \theta\}-\frac{1}{z+z_o}+\mathrm{e}^{2 \mathrm{i} \theta}\frac{\overline{z}+\overline{z_o}}{(z+z_o)^2} , \\
    & L_2(z, z_o; \theta)=-\overline{\beta} U\{J(z, z_o)+J(z, \overline{z_o}); \theta\}-\beta U\{(\overline{z_o}-z_o)G(z, \overline{z_o}); \theta\}-\frac{1}{\overline{z}+\overline{z_o}}-\mathrm{e}^{2 \mathrm{i} \theta}\frac{1}{z+z_o}, \\
    & U\{J(z, z_o); \theta\}= -J(z, z_o)+J(\overline{z},z_o)+\mathrm{e}^{2 \mathrm{i} \theta}(2 J(z,z_o)+(z-\overline{z})\frac{\partial J(z, z_o)}{\partial z}))
    \end{aligned}
    \label{eq:Lfunction}
\end{equation}
with $\beta=(1+\mathrm{i} \mu)/2$ and
\begin{equation}
\begin{aligned}
    J(z,z_o)&=\left[1-\frac{z}{z_o}\frac{(z_o^2-c^2)^{1/2}}{(z^2-c^2)^{1/2}}\right]\frac{z_o}{z^2-z_o^2}\\
    G(z,z_o)&=\frac{\partial J }{\partial z_o}
    =\left[1-\frac{z}{z_o}\frac{(z_o^2-c^2)^{1/2}}{(z^2-c^2)^{1/2}}\right]\frac{2 z_o^2}{z^2-z_o^2}+\left[1-\frac{z z_o}{(z_o^2-c^2)^{1/2}(z^2-c^2)^{1/2}}\right]\frac{1}{z^2-z_o^2}
\end{aligned}
\end{equation}

$S(z, \theta)$ represents the stresses at $z$ ( along the crack surface which shows an inclination angle of $\theta$ with respect to the sliding crack) resulting from a unit stress along the sliding crack $P^\prime P$.
 \begin{equation}
\begin{aligned}
    S(z,\theta)=\frac{1}{2} \mathrm{i} \{\frac{\overline{z}}{(\overline{z}^2-c^2)^{1/2}}-\frac{z}{(z^2-c^2)^{1/2}}+\mathrm{e}^{2 \mathrm{i} \theta} \left[ \frac{(\overline{z}-z)c^2}{(z^2-c^2)^{3/2}}+\frac{2z}{(z^2-c^2)^{1/2}}-2\right]\}
    \end{aligned}
    \label{eq:Sfunction}
\end{equation}
Note that both $L$ and $S$ functions are generalized functions. They may apply on the wing cracks or plastic zones, depending on which crack surface the parameters ($z$, $z_o$, $\alpha$, and $\theta$) are taken from.

$\sigma_{\hat{y}}^{\infty}$ and $\tau_{\hat{x}\hat{y}}^{\infty}$, and  $(\sigma_{\hat{\theta}}^{\infty}+\mathrm{i} \tau_{\hat{r} \hat{\theta}}^{\infty})_{\mathrm{t}}$  are respectively the projection of the far field stresses along the main sliding crack and the tensile wing crack.
\begin{equation}
    \begin{aligned}
    & \sigma_{\hat{y}}^{\infty}=\frac{\sigma_1+\sigma_2}{2}-\frac{\sigma_1-\sigma_2}{2} \cos 2 \gamma, \quad
\tau_{\hat{x}\hat{y}}^{\infty}=\frac{\sigma_1-\sigma_2}{2} \sin 2 \gamma \\
    & (\sigma_{\hat{\theta}}^{\infty}+\mathrm{i} \tau_{\hat{r} \hat{\theta}}^{\infty})_{\mathrm{t}}=\frac{\sigma_1+\sigma_2}{2} -\frac{\sigma_1-\sigma_2}{2} \cos 2 (\gamma-\theta)+\mathrm{i} \frac{\sigma_1-\sigma_2}{2} \sin 2 (\gamma-\theta)
    \end{aligned}
\end{equation}

\paragraph{For plastic zones}
Shear stresses are bounded along $PR$ and $P^{\prime}R^{\prime}$ with the constant yielding strength in shear $\tau_{\mathrm{y}}$. 
The singular integral form of Eq.~\eqref{eq:governingplastic} thus writes
\begin{equation}
\begin{aligned}
    \int_0^{\ell_{\mathrm{p}}} \frac{2 \overline{\alpha_{\mathrm{p}}}}{u-t}\text{d}t+\int_0^{\ell_{\mathrm{p}}} L(c+u, c+t; \alpha_{\mathrm{p}}, 0) \text{d}t+\\
    \mathrm{i} \, \text{Im}\left[\int_0^{\ell_{\mathrm{t}}} \alpha_{\mathrm{t}} T_1(u,r)\text{d}r+\int_0^{\ell_{\mathrm{t}}}\overline{\alpha_{\mathrm{t}}}T_2(u,r)\text{d}r+ \int_0^{\ell_{\mathrm{t}}}L(c+u, c+r\, \mathrm{e}^{\mathrm{i} \theta}; \alpha_{\mathrm{t}}, 0) \text{d}r\right]+\\
    \mathrm{i} \, \text{Im}\left[ 
    %\Phi_p^{\infty \prime}+
    (\tau_{\hat{x}\hat{y}}^{\infty}-\mu \sigma_{\hat{y}}^{\infty}+\tau_{\mathrm{c}}) S(c+u, 0) +(\sigma_{\hat{\theta}}^{\infty}+\mathrm{i} \tau_{\hat{r} \hat{\theta}}^{\infty})_{\mathrm{p}}\right]= -\mathrm{i} \tau_{\mathrm{y}}
\end{aligned}
\label{eq:Geqns3}
\end{equation}
with
\begin{equation}
        T_1(u, r)=\frac{1}{u-r\, \mathrm{e}^{\mathrm{i} \theta}}-\frac{u-r\, \mathrm{e}^{-\mathrm{i} \theta}}{(u-r\, \mathrm{e}^{\mathrm{i} \theta})^2}, \quad T_2(u, r)=\frac{1}{u-r\, \mathrm{e}^{-\mathrm{i} \theta}}+\frac{1}{u-r\, \mathrm{e}^{\mathrm{i} \theta}}
\end{equation}

Again the detailed expressions of $S$ and $L$ are respectively given in Eq.~\eqref{eq:Lfunction} and Eq.~\eqref{eq:Sfunction}.
$(\sigma_{\hat{\theta}}^{\infty}+\mathrm{i} \tau_{\hat{r} \hat{\theta}}^{\infty})_{\mathrm{p}}$ is the projection of the far field stresses on the plastic zone collinear with the sliding crack.
\begin{equation}
    (\sigma_{\hat{\theta}}^{\infty}+\mathrm{i} \tau_{\hat{r} \hat{\theta}}^{\infty})_{\mathrm{p}}=\frac{\sigma_1+\sigma_2}{2} -\frac{\sigma_1-\sigma_2}{2} \cos 2 \gamma+\mathrm{i} \frac{\sigma_1-\sigma_2}{2} \sin 2 \gamma
    \label{eq:Geqns4}
\end{equation}

The zero stress singularity at the end of the plastic zone 
 Eq.~(\ref{eq:boundaryplastic}) can be also expressed in function of the dislocation density $\alpha_\mathrm{p}$:
\begin{equation}
    \mathrm{i} K_{\mathrm{II}}^R=\lim_{t \to \ell_{\mathrm{p}}} (2 \pi)^{3/2} \overline{\alpha_{\mathrm{p}}} (\ell_{\mathrm{p}}-t)^{1/2}=0
\end{equation}

Solving the discolation densities $\alpha_\mathrm{t}$ from the governing equations \eqref{eq:Geqns1}, \eqref{eq:Geqns2}, \eqref{eq:Geqns3} and \eqref{eq:Geqns4}, we obtain the stress intensity factor at the tips $Q$ and $Q^{\prime}$ of the wing cracks:
\begin{equation}
    K_{\mathrm{I}}+\mathrm{i} K_{\mathrm{II}}=(2 \pi)^{3/2} \lim_{r \to \ell_{\mathrm{t}}} (\ell_{\mathrm{t}}-r)^{1/2} \overline{\alpha_{\mathrm{t}}} \mathrm{e}^{\mathrm{i} \theta}
\label{eq:stressintensity3}
\end{equation}
Most of the mathematical formulation presented here is taken from \cite{NeHo1982,HoNe1985,HoNe1986} among which inconsistencies exist and have been mostly reconciled in \cite{HoNe1986}. Note that a typo exists in the definition of $L_2$ in all three papers. We have now corrected the typo in  $L_2$ using the expression from \cite{Nema1983} and reorganized the formulation hoping to reconcile all inconsistencies present in previous studies \cite{NeHo1982,Nema1983,HoNe1985,HoNe1986}. We recall here that compression is viewed negative in the above formulations \cite{NeHo1982,Nema1983,HoNe1985,HoNe1986}, while in rock mechanics, compression is often viewed as positive (as in the main text). 

\section{Gauss-Chebyshev quadrature and barycentric interpolation techniques \label{ax:Numerical}}
Gauss-Chebyshev quadrature method is a classical technique for the solution of singular integral equation arising in fracture mechanics. The method makes use of a primary and a complimentary sets of nodes, discretizing a fracture interval of $(-1,1)$. In this paper, we adopt the first type of quadrature $T_k$ along the kinked crack and the plastic zone, where the primary sets of nodes $s^{\prime}_i, u^{\prime}_i$ and the complimentary sets of nodes $r^{\prime}_j, t^{\prime}_j$ are given as follows
\begin{equation}
    s^{\prime}_i, u^{\prime}_i=\cos\left(\frac{\pi i}{n}\right), \quad r^{\prime}_j, t^{\prime}_j=\cos\left(\frac{\pi (j-1/2)}{n}\right), \quad i=1, ..., n-1,\quad j=1, ..., n 
\end{equation}
This choice of the first type of quadrature embeds the square-root singularity of linear elastic fracture mechanics at both ends of the fracture domain. The dislocation density $\text{d} \Omega/\text{d} r^{\prime}$ of the kinked crack is expressed as:
\begin{equation}
    \frac{\text{d} \Omega}{\text{d} r^{\prime}}=\omega(r^{\prime})F(r^{\prime}), \quad  w(r^{\prime})=\frac{1}{\sqrt{(1-r^{\prime}_j)(1+r^{\prime}_j)}}
\label{eq:Fdefinition}
\end{equation}
where $\Omega$ represents the dimensionless crack opening/slip or dislocation, $\omega(r^{\prime})$ a weight function with the required tip singularity and $F(r^{\prime})$ an unknown non-singular function. The dislocation density of the plastic zone can be obtained by replacing $r^{\prime},\,s^{\prime}$ with $t^{\prime},\,u^{\prime}$. 

In the following, we present some important operators used to discretize the governing equations. We use $s^{\prime}$ and $r^{\prime}$ (corresponding to the kinked crack) for simplicity, which however can be easily applied to $u^{\prime}$ and $t^{\prime}$ (corresponding to the plastic zone).

The finite Hilbert transform is evaluated on the complimentary $\mathbf{s^\prime}$-set of the nodes using representation of the dislocation density \eqref{eq:Fdefinition} on the $\mathbf{r^\prime}$-set, ${F(r^{\prime}_j)}$. This results in the following:
\begin{equation}
    \frac{1}{\pi}\int_{-1}^1\frac{\partial \Omega}{\partial r^{\prime}}\frac{1}{s^{\prime}-r^{\prime}}\text{d}r^{\prime}\approx \sum_{j=1}^n H_{ij} F(r^{\prime}_j), \quad H_{ij}=\frac{1}{n} \sum_{j=1}^n \frac{1}{s^{\prime}_i-r^{\prime}_j}
\end{equation}

The unknown function $F$ representing the unknown dislocation density \eqref{eq:Fdefinition} can be extrapolated from the Gauss-Chebyshev nodes to the fracture tip:
\begin{equation}
    F(1)\approx \sum_{j=1}^n Q_j F(r^{\prime}_j), \quad Q_j=-(-1)^j \frac{\cot((\arccos{r^{\prime}_j})/2)}{n}
\end{equation}
\begin{equation}
    F(-1)\approx \sum_{j=1}^n P_j F(r^{\prime}_j), \quad P_j=(-1)^j \frac{\tan((\arccos{r^{\prime}_j})/2)}{n}
\end{equation}

With a zero-dislocation boundary condition at the fracture tip $\Omega(1)=0$, we can also evaluate the corresponding crack opening and slip by integrating the dislocation density along the crack surface.

\begin{equation}
    \Omega(s^{\prime}_i)=\int_{1}^{s^\prime_i}\frac{\partial \Omega}{\partial r^\prime}\mathrm{d}r^\prime \approx \sum_{j=1}^n S^{*}_{ij} F(r^{\prime}_j), \quad S^{*}_{ij}=\sum_{k=0}^{n-1} [\Phi_k(s^{\prime}_i)-\Phi_k(1)]B_{kj}
\end{equation}

\begin{equation}
    \Omega(-1)=\int_{1}^{-1}\frac{\partial \Omega}{\partial r^\prime}\mathrm{d}r^\prime \approx  \sum_{j=1}^n S^{**}_{j} F(r^{\prime}_j), \quad S^{**}_{j}=\sum_{k=0}^{n-1} [\Phi_k(-1)-\Phi_k(1)]B_{kj}
\end{equation}

The integration of the slip or opening along the crack surface can be further obtained using partial differentiation: 
\begin{equation}
    \begin{aligned}
       \int_{-1}^{1} \Omega (s^{\prime}) \text{d}s^{\prime}&=\left.s^{\prime} \Omega(s^{\prime})\right|_{-1}^{1}-\int_{-1}^1 r^{\prime} \frac{\partial \Omega}{\partial r^{\prime}}\text{d}r^{\prime}
       =\Omega(-1)-\sum_{j=1}^n S^{**}_j (r^{\prime}_j F(r^\prime_j))
    \end{aligned}
\end{equation}
We refer the reader to \cite{ViGa2017,LiLG2019} for the detailed expressions of all above operators.

\section{Discretization \label{ax:Discretization}}
We map all the coordinates $s, r \in (0, \ell_{\mathrm{t}})$ along the wing crack and the plastic zone $u, t \in (0, \ell_{\mathrm{p}})$, onto the range of $(-1,1)$.
\begin{equation}
\begin{aligned}
    s=\frac{s^{\prime}+1}{2} \ell_{\mathrm{t}},\quad r=\frac{t^{\prime}+1}{2} \ell_{\mathrm{t}} \quad s, r \in (0, \ell_{\mathrm{t}}), \quad s^{\prime}, r^{\prime} \in (-1,1)\\
    u=\frac{u^{\prime}+1}{2} \ell_{\mathrm{p}},\quad t=\frac{t^{\prime}+1}{2} \ell_{\mathrm{p}} \quad u, t \in (0, \ell_{\mathrm{p}}), \quad u^{\prime}, t^{\prime} \in (-1,1)
\end{aligned}
\end{equation} 
We scale the crack length and width/slip with different characteristic length scales.
\begin{equation}
    \ell_{\mathrm{t}}=\ell_{\mathrm{t}}^{\prime}c, \quad \ell_{\mathrm{p}}=\ell_{\mathrm{p}}^{\prime} c, \quad [v] \sim \Omega \frac{\sigma_1}{E^{\prime}}c
\end{equation}
where $\ell_{\mathrm{t}}^{\prime}, \ell_{\mathrm{p}}^{\prime}$ are dimensionless length of the wing crack and plastic zone. $\Omega$ represents the dimensionless crack opening/slip or dislocation.

% Give the detailed expressions of w_m w_j

% The quadrature weight becomes
% \begin{equation}
%     w_j=\pi \frac{1}{n}
% \end{equation}

\paragraph{For wing cracks}
We discretize the boundary condition Eq.~\eqref{eq:Geqns1} representing a zero dislocation normal to the pre-existing flaw and the stress-free condition Eq.~\eqref{eq:Geqns2} as follows:
\begin{equation}
    \sum_{j=1}^{n}\left(S^{**}_{j}F_r(r^{\prime}_j) \sin \theta +S^{**}_{j} F_{\theta}(r^{\prime}_j) \cos \theta\right)=0
    \label{eq:Deqns1}
\end{equation}

\begin{equation}
    \begin{aligned}
    \frac{1}{2} \frac{1}{\ell_{\mathrm{t}}^{\prime}} \sum_{j=1}^n H_{ij} \left(F_{\theta}(r^{\prime}_j)+\mathrm{i} F_r(r^{\prime}_j)\right)+\\
    \frac{1}{8 \pi}  \sum_{j=1}^n  \left( \frac{\pi}{n}\right) \mathcal{L}_{1,ij}^{\mathrm{tt}} \left((F_r(r^{\prime}_j)\sin \theta+F_{\theta}(r^{\prime}_j)\cos \theta)-\mathrm{i}(F_r(r^{\prime}_j)\cos\theta-F_{\theta}(r^{\prime}_j)\sin \theta) \right) +\\ 
    \frac{1}{8 \pi}  \sum_{j=1}^n \left( \frac{\pi}{n}\right) \mathcal{L}_{2,ij}^{\mathrm{tt}} \left((F_r(r^{\prime}_j)\sin \theta+F_{\theta}(r^{\prime}_j)\cos \theta)+\mathrm{i}(F_r(r^{\prime}_j)\cos\theta)-F_{\theta}(r^{\prime}_j)\sin \theta\right)+\\
      \frac{1}{8\pi}\sum_{m=1}^n \left(\frac{\pi}{n}\right) \mathcal{L}_{1,im}^{\mathrm{tp}} (-\mathrm{i} F_p(t_m^{\prime}))+\frac{1}{8\pi}\sum_{m=1}^n \left(\frac{\pi}{n}\right) \mathcal{L}_{2,im}^{\mathrm{tp}} (\mathrm{i} F_p(t_m^{\prime}))\\
      \frac{1}{8\pi} \sum_{m=1}^n \left(\frac{\pi}{n}\right)  \mathcal{K}_{1,im}^{\mathrm{tp}}(-\mathrm{i} F_p(t^{\prime}_m))+
      \frac{1}{8 \pi}\sum_{m=1}^n  \left(\frac{\pi}{n}\right) \mathcal{K}_{2, im}^{\mathrm{tp}}(\mathrm{i} F_p(t_m^{\prime}))+\\
       \mathcal{S}_{i}^{\mathrm{t}} \left(\frac{1}{2}(1-\frac{\sigma_2}{\sigma_1}) (\sin 2 \gamma +\mu  \cos 2 \gamma)-\frac{1}{2} \mu (1+\frac{\sigma_2}{\sigma_1}) +\frac{\tau_{\mathrm{c}}}{\sigma_1}\right)+\\
      \frac{1}{2} (1+\frac{\sigma_2}{\sigma_1})-\frac{1}{2}(1-\frac{\sigma_2}{\sigma_1}) \cos 2 (\gamma-\theta)+\mathrm{i} \left(\frac{1}{2}(1-\frac{\sigma_2}{\sigma_1}) \sin 2 (\gamma-\theta)\right)=0\\
\end{aligned}
\label{eq:Deqns2}
\end{equation}
with 
% should distinguish between m and k between i and j
\begin{equation}
\begin{aligned}
    &\mathcal{L}_{1, ij}^{\mathrm{tt}}=L_1(1+\frac{s_i^{\prime}+1}{2}\ell_{\mathrm{t}}^{\prime} \mathrm{e}^{\mathrm{i} \theta},1+\frac{r_j^{\prime}+1}{2}\ell_{\mathrm{t}}^{\prime} \mathrm{e}^{\mathrm{i} \theta}; \theta, c=1), \\
        &\mathcal{L}_{2, ij}^{\mathrm{tt}}=L_2(1+\frac{s_i^{\prime}+1}{2}\ell_{\mathrm{t}}^{\prime} \mathrm{e}^{\mathrm{i} \theta},1+\frac{r_j^{\prime}+1}{2}\ell_{\mathrm{t}}^{\prime} \mathrm{e}^{\mathrm{i} \theta}; \theta, c=1),\\
        &\mathcal{L}_{1, im}^{\mathrm{tp}}=L_1(1+\frac{s_i^{\prime}+1}{2}\ell_{\mathrm{t}}^{\prime} \mathrm{e}^{\mathrm{i} \theta},1+\frac{t_m^{\prime}+1}{2}\ell_{\mathrm{p}}^{\prime}; \theta , c=1), \\
        &\mathcal{L}_{2, im}^{\mathrm{tp}}=L_2(1+\frac{s_i^{\prime}+1}{2}\ell_{\mathrm{t}}^{\prime} \mathrm{e}^{\mathrm{i} \theta},1+\frac{t_m^{\prime}+1}{2}\ell_{\mathrm{p}}^{\prime}; \theta, c=1),\\
&\mathcal{S}_{ i}^{\mathrm{t}}=S(1+\frac{s_i^{\prime}+1}{2}\ell_{\mathrm{t}}^{\prime} \mathrm{e}^{\mathrm{i} \theta}, \theta; c=1),\\
    &\mathcal{K}_{1,im}^{\mathrm{tp}}=K_1(\frac{s_i^{\prime}+1}{2} \ell_{\mathrm{t}}^{\prime}, \frac{t_m^{\prime}+1}{2} \ell_{\mathrm{p}}^{\prime}), \\
    &\mathcal{K}_{2,im}^{\mathrm{tp}}=K_2(\frac{s_i^{\prime}+1}{2} \ell_{\mathrm{t}}^{\prime}, \frac{t_m^{\prime}+1}{2} \ell_{\mathrm{p}}^{\prime})
\end{aligned}
\end{equation}
where $F_r$ and $F_{\theta}$ are unknowns associated with the mode I (opening mode) and mode II (shear mode) dislocation density on the wing cracks and $F_p$ are unknowns associated with the mode II dislocation density in the plastic zone. $H_{ij}$ and $S_j^{**}$ are operators reported in Appendix.~\ref{ax:Numerical}.

\paragraph{For plastic zones}
The discretized form of the bounded yielding shear stress condition Eq.~\eqref{eq:Geqns3} becomes 
\begin{equation}
\begin{aligned}
    \frac{1}{2} \frac{1}{\ell_{\mathrm{p}}^{\prime}}\sum_{m=1}^n  H_{km} (\mathrm{i} F_p(t_m^{\prime})) +\\
    \frac{1}{8\pi} \sum_{m=1}^n \left(\frac{\pi}{n}\right)  \mathcal{L}_{1, km}^{\mathrm{pp}}( -\mathrm{i} F_p(t_m^{\prime}))+ \frac{1}{8\pi} \sum_{m=1}^n \left(\frac{\pi}{n}\right)  \mathcal{L}_{2, km}^{\mathrm{pp}}( \mathrm{i} F_p(t_m^{\prime}))+\\
    \mathrm{i} \frac{1}{8 \pi} \text{Im}[\sum_{j=1}^n \left(\frac{\pi}{n}\right) \mathcal{T}_{1, kj}^{\mathrm{pt}} \left((F_r(r^{\prime}_j)\sin \theta+F_{\theta}(r^{\prime}_j)\cos \theta)-\mathrm{i}(F_r(r^{\prime}_j)\cos\theta-F_{\theta}(r^{\prime}_j)\sin \theta) \right)]+\\
    \mathrm{i}  \frac{1}{8\pi} \text{Im}[\sum_{j=1}^n \left(\frac{\pi}{n}\right) \mathcal{T}_{2, kj}^{\mathrm{pt}} \left((F_r(r^{\prime}_j)\sin \theta+F_{\theta}(r^{\prime}_j)\cos \theta)+\mathrm{i}(F_r(r^{\prime}_j)\cos\theta)-F_{\theta}(r^{\prime}_j)\sin \theta\right)]+\\
    \mathrm{i} \frac{1}{8 \pi} \text{Im}[ \sum_{j=1}^n \left(\frac{\pi}{n}\right) \mathcal{L}_{1, kj}^{\mathrm{pt}}  \left((F_r(r^{\prime}_j)\sin \theta+F_{\theta}(r^{\prime}_j)\cos \theta)-\mathrm{i}(F_r(r^{\prime}_j)\cos\theta-F_{\theta}(r^{\prime}_j)\sin \theta) \right)]+\\
    \mathrm{i} \frac{1}{8 \pi} \text{Im}[ \sum_{j=1}^n \left(\frac{\pi}{n}\right) \mathcal{L}_{2, kj}^{\mathrm{pt}} \left((F_r(r^{\prime}_j)\sin \theta+F_{\theta}(r^{\prime}_j)\cos \theta)+\mathrm{i}(F_r(r^{\prime}_j)\cos\theta-F_{\theta}(r^{\prime}_j)\sin \theta) \right)]+ \\
    \mathrm{i} \mathcal{S}_{k}^{\mathrm{p}} \left(\frac{1}{2}(1-\frac{\sigma_2}{\sigma_1}) (\sin 2 \gamma +\mu  \cos 2 \gamma)-\frac{1}{2} \mu (1+\frac{\sigma_2}{\sigma_1}) +\frac{\tau_{\mathrm{c}}}{\sigma_1}\right)+
      \mathrm{i} \left(\frac{1}{2}(1-\frac{\sigma_2}{\sigma_1}) \sin 2 \gamma \right)+ \mathrm{i} \frac{\tau_{\mathrm{y}}}{\sigma_1}=0
\end{aligned}
\label{eq:Deqns3}
\end{equation}
with
\begin{equation}
\begin{aligned}
    &\mathcal{L}_{1, km}^{\mathrm{pp}}=L_{1}(1+\frac{u_k^{\prime}+1}{2}\ell_{\mathrm{p}}^{\prime} ,1+\frac{t_m^{\prime}+1}{2}\ell_{\mathrm{p}}^{\prime}; 0, c=1), \\
        &\mathcal{L}_{2, km}^{\mathrm{pp}}=L_{2}(1+\frac{u_k^{\prime}+1}{2}\ell_{\mathrm{p}}^{\prime} ,1+\frac{t_m^{\prime}+1}{2}\ell_{\mathrm{p}}^{\prime}; 0, c=1),\\
        &\mathcal{L}_{1, kj}^{\mathrm{pt}}=L_{1}(1+\frac{u_k^{\prime}+1}{2}\ell_{\mathrm{p}}^{\prime} ,1+\frac{r_j^{\prime}+1}{2}\ell_{\mathrm{t}}^{\prime} \mathrm{e}^{\mathrm{i} \theta}; 0, c=1), \\
        &\mathcal{L}_{2, kj}^{\mathrm{pt}}=L_{2}(1+\frac{u_k^{\prime}+1}{2}\ell_{\mathrm{p}}^{\prime} ,1+\frac{r_j^{\prime}+1}{2}\ell_{\mathrm{t}}^{\prime} \mathrm{e}^{\mathrm{i} \theta}; 0, c=1),\\
        &\mathcal{S}_{k}^{\mathrm{p}}=S(1+\frac{u^{\prime}_k+1}{2} \ell_{\mathrm{p}}^{\prime}, 0; c=1),\\
        &\mathcal{T}_{1, kj}^{\mathrm{pt}}=T_1(\frac{u_k^{\prime}+1}{2} \ell_{\mathrm{p}}^{\prime}, \frac{r_j^{\prime}+1}{2} \ell_{\mathrm{t}}^{\prime}), \\
        &\mathcal{T}_{2, kj}^{\mathrm{pt}}=T_2(\frac{u_k^{\prime}+1}{2} \ell_{\mathrm{p}}^{\prime}, \frac{r_j^{\prime}+1}{2} \ell_{\mathrm{t}}^{\prime})
\end{aligned}
\end{equation}

The zero-singularity at the tips of the plastic zones Eq.~(\ref{eq:Geqns4}) writes
\begin{equation}
\begin{aligned}
    \mathrm{i} \frac{1}{\tau_{\mathrm{y}} \sqrt{\pi c}} K_{\mathrm{II}}^R=\frac{1}{2\sqrt{2}} \frac{\sigma_1}{\tau_{\mathrm{y}}} (\ell_{\mathrm{p}}^{\prime})^{-1/2} \left(\mathrm{i} F_p(1) \right)=0, \quad \sum_{m=1}^n Q_m F_p(t_m^{\prime})=0
\end{aligned}
\label{eq:Deqns4}
\end{equation}
where $Q_m$ indicates the interpolation vector reported in Appendix.~\ref{ax:Numerical}.

\paragraph{Additional constraints at the kink}
The problem turns into a set of equations discritized by the first type of Gauss-Chebysheve quadrature $T_k$: there are $(3n+1)$ unknowns ($n$ unknowns in $F_r$, $F_{\theta}$ and $F_p$ respectively and one unknown related to the maximum stress in compression $\sigma_1/\tau_{\mathrm{y}}$), and $3(n-1)$ equations of elasticity \eqref{eq:Deqns2}, \eqref{eq:Deqns3}, and two boundary conditions \eqref{eq:Deqns1}, \eqref{eq:Deqns4}. One needs to add two more equations to solve the problem with the same number of unknowns and equations. In this paper, we add two extra constraints at the kink points $P$ and $P^{\prime}$, which are not directly reported in previous studies \cite{NeHo1982,HoNe1985,Nema1983,HoNe1986}.  

\cite{HeHu1989} points out that a weaker singularity exists at the kink where the most singular stresses in the vicinity of the kink follows the form $\sigma \sim r^{-\eta}$, where $\eta$ is a real number and $\eta<1/2$. %This means that we have zero singularity at the kink point for mode II and a singularity smaller than $1/2$ for mode I. We therefore have
The dislocation density can be therefore written as 
\begin{equation}
    \frac{\text{d}\Omega}{\text{d}r}=(1+r^{\prime})^{-\eta}(1-r^{\prime})^{-1/2}F_T(r^{\prime}), \quad \eta<1/2
\end{equation}
where $(1+r^{\prime})^{-\eta}(1-r^{\prime})^{-1/2}$ are weight functions associated with the weak stress singularity at the kink and $F_T$ an unknown non-singular function. Interestingly, the dislocation density is also defined in Eq.~\eqref{eq:Fdefinition} using a different weight function which represents a square-root stress singularity associated with the first type of Gauss-Chebyshev quadrature $T_k$. By comparing these two definitions, we obtain the relation between the two non-singular functions at the kink ($r^{\prime}=-1$):
\begin{equation}
    F(-1)=F_T(-1) \lim_{r^{\prime} \to -1} (1+r^{\prime})^{1/2-\eta}=0
\end{equation}
This imposes two more constraints on the kink related to the tensile wing cracks:
\begin{equation}
     F_{\theta}(-1)=0, \quad \sum_{j=1}^n P_j F_{\theta}(r^{\prime}_j)=0
     \label{eq:KinkCondition1}
\end{equation}
\begin{equation}
     F_{r}(-1)=0, \quad \sum_{j=1}^n P_j F_r(r^{\prime}_j)=0   
\label{eq:KinkCondition2}
\end{equation}
Likewise, accounting for the constant yielding shear stress inside the plastic zone: the zero stress singularity condition should apply on both ends of the plastic zone $R$, $R^{\prime}$ and the kink points $P$, $P^{\prime}$.  Such a constraint is however only imposed at $R$, $R^{\prime}$ in \cite{HoNe1986}. We therefore impose the constraint on the other ends of the plastic zone $P$, $P^{\prime}$ via the dislocation density $\mathrm{d}\Omega/\mathrm{d}t^\prime=\omega (t^\prime)F_p (t^\prime)=0$ \eqref{eq:Fdefinition}. This leads to  
\begin{equation}
    F_p(-1)=0, \quad \sum_{m=1}^n P_m F_p(t_m^{\prime})=0
\label{eq:KinkCondition3}
\end{equation}

In this paper, we use Eq.~(\ref{eq:KinkCondition2}) and Eq.~(\ref{eq:KinkCondition3}) as the two additional equations to solve the problem. Eq.~(\ref{eq:KinkCondition1}) and Eq.~(\ref{eq:KinkCondition1}) act as equivalent conditions: the numerical simulation gives the same results if we choose Eq.~(\ref{eq:KinkCondition1}) over Eq.~(\ref{eq:KinkCondition2}) or vice versa.

\begin{figure}
    \centering
    \begin{tabular}{ccc}
          \includegraphics[height=0.36 \linewidth]{ 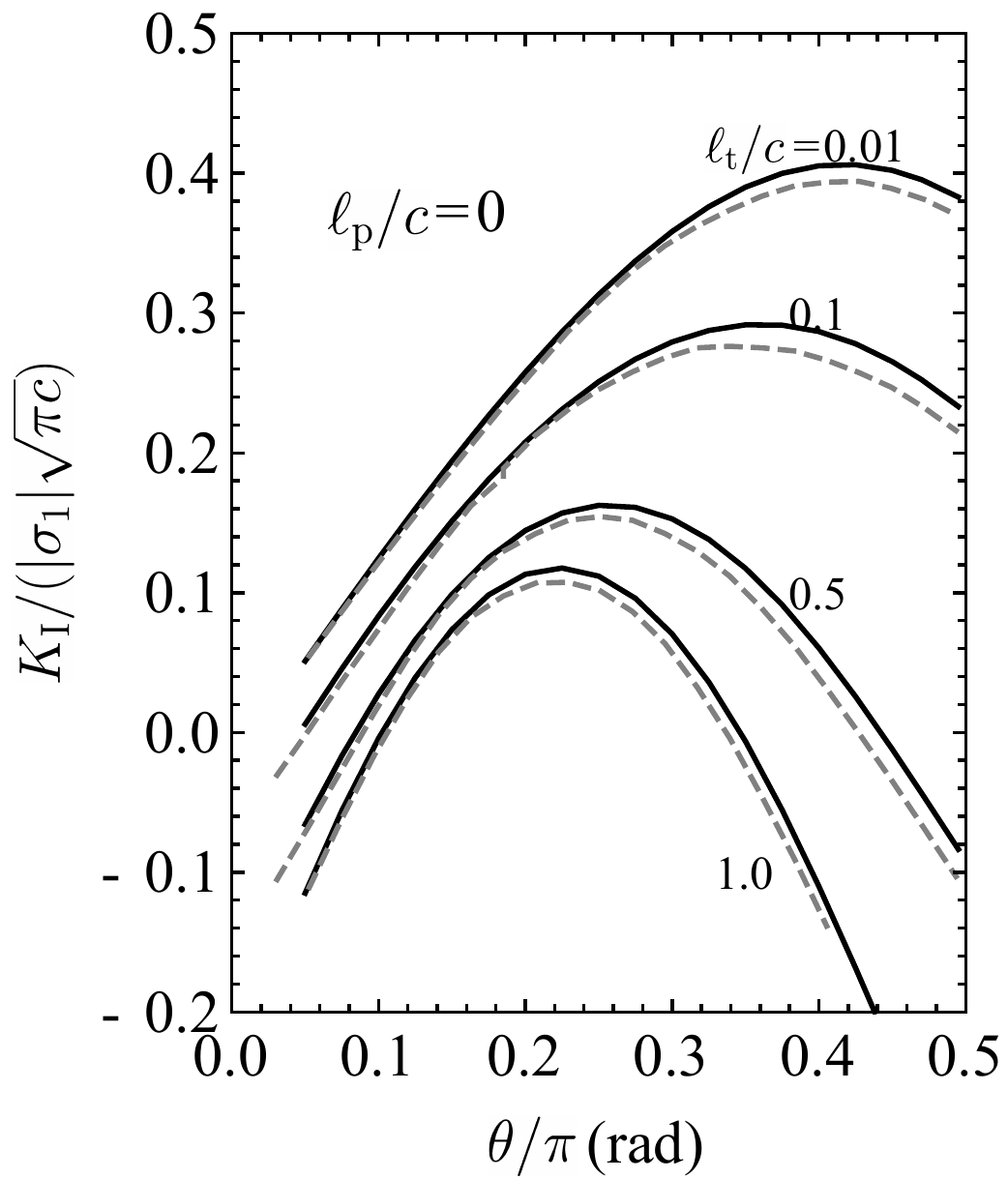}&
          \includegraphics[height=0.36 \linewidth]{ 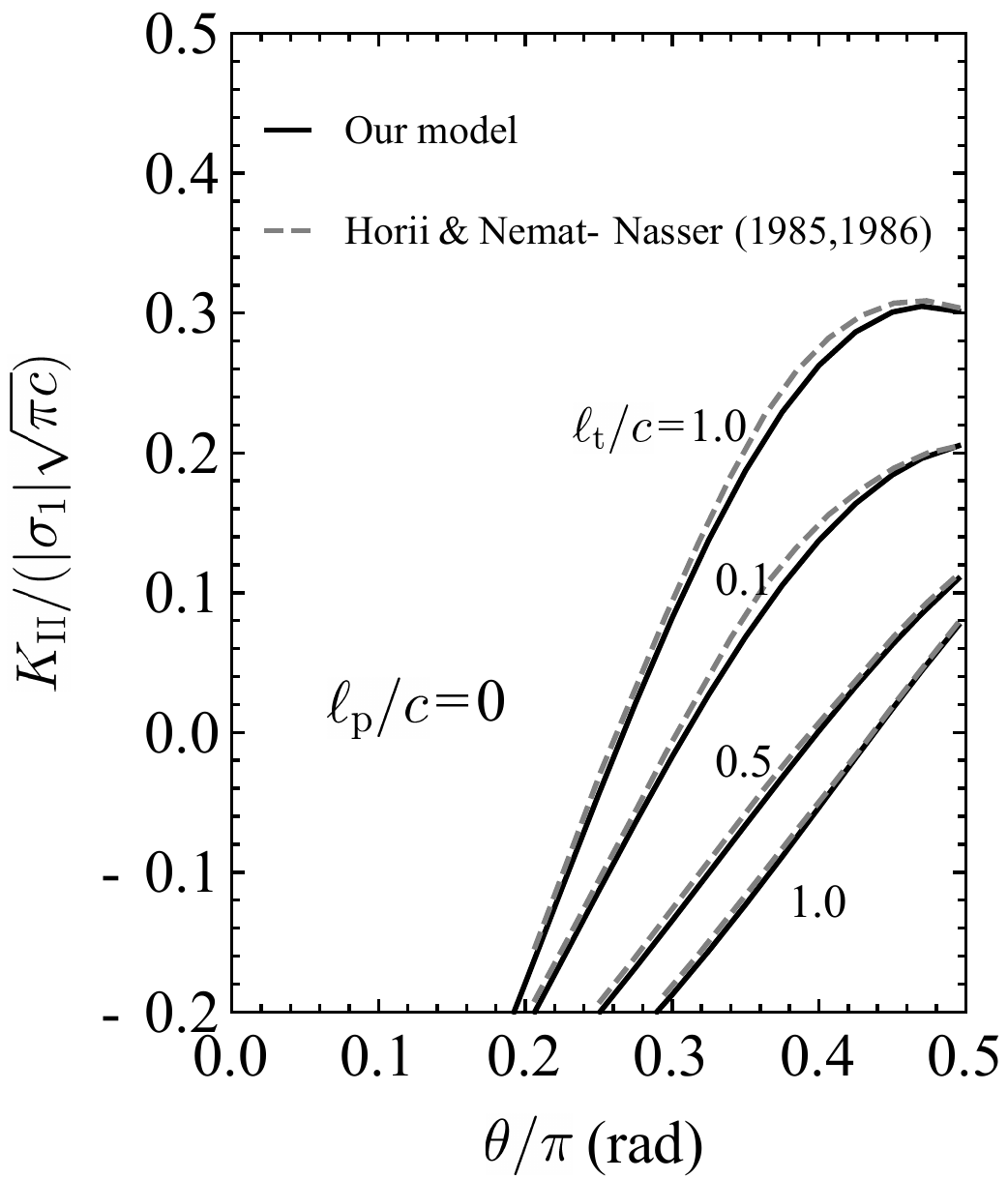}&
           \includegraphics[height=0.36 \linewidth]{ 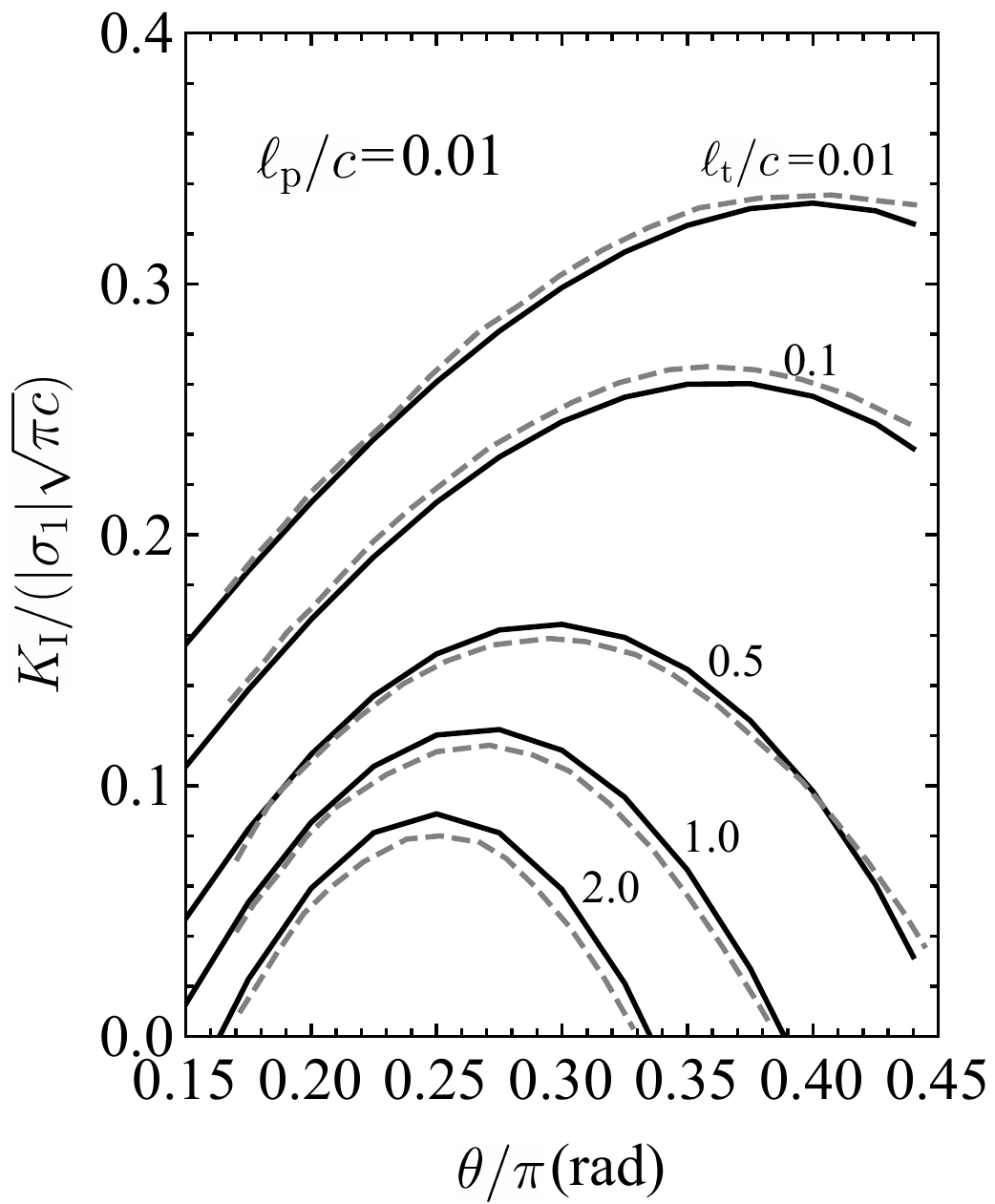}\\
         (a) & (b) & (c)\\ 
    \end{tabular}
    \caption{Benchmark of the numerical code: stress intensity factors at the tip $Q$ of the straight wing crack, as functions of the crack orientation angle $\gamma$ for indicated crack length; (a) and (b): $\ell_{\mathrm{p}}/c= 0$,\,$\tau_{\mathrm{c}}= 0, \, \mu= 0.3, \,\sigma_2/\sigma_1= 0$, and $\gamma= \pi/5$. (c): $\ell_{\mathrm{p}}/c= 0.01$, $\tau_{\mathrm{c}}= 0,\, \mu= 0.4,\, \sigma_2/\sigma_1= 0$, and $\gamma= \pi/4$. }
    \label{fig:Benchmark}
\end{figure}

\paragraph{Stress intensity factors}
Solving $F_r$, $F_{\theta}$, $F_p$ and $\sigma_1/\tau_{\mathrm{y}}$ subject to equations \eqref{eq:Deqns1}, \eqref{eq:Deqns2}, \eqref{eq:Deqns3}, \eqref{eq:Deqns4}, \eqref{eq:KinkCondition2}, \eqref{eq:KinkCondition2}, we are able to calculate the stress intensity factors at the crack tips from 
\begin{equation}
\begin{aligned}
    \frac{1}{\tau_{\mathrm{y}} \sqrt{\pi c}}(K_{\mathrm{I}}+\mathrm{i} K_{\mathrm{II}})
    &=\frac{1}{2\sqrt{2}}\frac{\sigma_1}{\tau_{\mathrm{y}}}\left(\ell_{\mathrm{t}}^{\prime}\right)^{-1/2}  \left( F_{\theta}(1)+\mathrm{i}F_r(1)\right)\\
    &=\frac{1}{2\sqrt{2}}\frac{\sigma_1}{\tau_{\mathrm{y}}}\left(\ell_{\mathrm{t}}^{\prime}\right)^{-1/2} \sum_{m=1}^n \left( Q_j F_{\theta}(r_j^{\prime})+\mathrm{i} Q_j F_r(r_j^{\prime})\right)\\
    \frac{1}{|\sigma_1| \sqrt{\pi c}}(K_{\mathrm{I}}+\mathrm{i} K_{\mathrm{II}})&=-\frac{1}{2\sqrt{2}}\left(\ell_{\mathrm{t}}^{\prime}\right)^{-1/2}  \left( F_{\theta}(1)+\mathrm{i}F_r(1)\right)\\
    &=-\frac{1}{2\sqrt{2}}\left(\ell_{\mathrm{t}}^{\prime}\right)^{-1/2} \sum_{m=1}^n \left(Q_j F_{\theta}(r_j^{\prime})+\mathrm{i} Q_j F_r(r_j^{\prime})\right)
\end{aligned}
\end{equation}
By estimating the stress intensity factors at the tips of wing cracks, we benchmark our numerical solver using the same configurations as those reported in \cite{HoNe1985,HoNe1986}. Our numerical results agree well with those in \cite{HoNe1985,HoNe1986} as shown in Fig.~\ref{fig:Benchmark}.

\section{Inelastic strain calculation\label{ax:Strain}}
In the following, we present the estimation of the inelastic strains based on the dislocation along the crack surfaces. Note that for $\sigma_1, \, \sigma_2$, compression is viewed negative, while for the inelastic strain, contraction is viewed positive.
The inelastic strain $\epsilon^{\mathrm{p}}$ consists of the deformation due to the sliding of the pre-existing flaw $\epsilon^{\mathrm{pc}}$, the dislocations in the plastic zones $\epsilon^{\mathrm{pp}}$ and tensile crack opening of the wings $\epsilon^{\mathrm{pt}}$:
\begin{equation}
\epsilon^{\mathrm{p}}=\epsilon^{\mathrm{pc}}+\epsilon^{\mathrm{pp}}+\epsilon^{\mathrm{pt}}
    \label{eq:Strain}
\end{equation}
\paragraph{Strains due to the sliding crack and the plastic zone}
Following Eq.~\eqref{eq:inelasticstrain}, in a local coordinate system $(\hat{x}, \hat{y})$ as shown in Fig.~\ref{fig:StrainRotation},
the strain due to the sliding crack writes
\begin{equation}
    \epsilon_{\hat{x}\hat{y}}^{\mathrm{pc}}=\epsilon_{\hat{y}\hat{x}}^{\mathrm{pc}}=
    %\frac{f}{c^2}\int_0^c [v_2] \text{d}x=\frac{f}{c^2} v_x c = f\frac{\delta_x}{c}=
    f \frac{\sigma_1}{E^{\prime}}\left(\Omega_{\theta}(-1)\sin{\theta}-\Omega_{r}(-1) \cos{\theta}\right), \quad \epsilon_{\hat{x}\hat{x}}^{\mathrm{pc}}=\epsilon_{\hat{y}\hat{y}}^{\mathrm{pc}}=0
\end{equation}
where $\Omega_{\theta}(-1)$ and $\Omega_r(-1)$ represent respectively the crack opening and shear displacement of the wing crack at the kink $P$.

Likewise, we obtain the strain due to the dislocation inside the plastic zone:

\begin{equation}
    \epsilon_{\hat{x}\hat{y}}^{\mathrm{pp}}=\epsilon_{\hat{y}\hat{x}}^{\mathrm{pp}}%=2 \frac{1}{2}\frac{f}{c^2}\int_0^{\ell_{\mathrm{p}}} -\delta_p \text{d}\ell
    %=\frac{N}{V} \frac{1}{2} \ell_{\mathrm{p}} \int_{-1}^{1} -\delta_p \text{d}u^{\prime}
    =-f \frac{1}{2} \frac{\ell_{\mathrm{p}}}{c} \frac{\sigma_1}{E^{\prime}} \left(\Omega_p(-1)-\int_{-1}^{1} u^{\prime} \frac{\partial \Omega_p}{\partial u^{\prime}} \text{d}u^{\prime}\right), \quad
    \epsilon_{\hat{x}\hat{x}}^{\mathrm{pp}}=\epsilon_{\hat{y}\hat{y}}^{\mathrm{pp}}=0
\end{equation}

In the global coordinate system $x$, $y$,
the inelastic strain due to the slip and dislocation along the slip line thus becomes 
\begin{equation}
\begin{aligned}
    &\epsilon_{yy}^{\mathrm{pp}}+\epsilon_{yy}^{\mathrm{pc}}=-\sin{2 \gamma}(\epsilon_{\hat{x}\hat{y}}^{\mathrm{pp}}+\epsilon_{\hat{x}\hat{y}}^{\mathrm{pc}}) \\
    &\epsilon_{yx}^{\mathrm{pp}}+\epsilon_{yx}^{\mathrm{pc}}=\epsilon_{xy}^{\mathrm{pp}}+\epsilon_{xy}^{\mathrm{pc}}=-\cos{2 \gamma}(\epsilon_{\hat{x}\hat{y}}^{\mathrm{pp}}+\epsilon_{\hat{x}\hat{y}}^{\mathrm{pc}})\\
    &\epsilon_{xx}^{\mathrm{pp}}+\epsilon_{xx}^{\mathrm{pc}}=\sin{2 \gamma}(\epsilon_{\hat{x}\hat{y}}^{\mathrm{pp}}+\epsilon_{\hat{x}\hat{y}}^{\mathrm{pc}})
\end{aligned}
\end{equation}

\paragraph{Strains due to the tensile wing cracks}

From Eq.~\eqref{eq:inelasticstrain}, we obtain the strains  expressed in the local coordinate system of $\hat{r}$, $\hat{\theta}$ as shown in Fig.~\ref{fig:StrainRotation}:
\begin{equation}
    \begin{aligned}
        &\epsilon_{\hat{\theta}\hat{\theta}}^{\mathrm{pt}}=2 \frac{1}{2} f \frac{\sigma_1}{E^{\prime}} \frac{\ell_{\mathrm{t}}}{c}\int_{-1}^{1} \Omega_{\theta} \text{d}r^{\prime}= f \frac{\sigma_1}{E^{\prime}} \frac{\ell_{\mathrm{t}}}{c}\left(\Omega_{\theta}(-1)-\int_{-1}^{1}r^{\prime} \frac{\partial \Omega_{\theta}}{\partial r^{\prime}} \text{d}r^{\prime}\right)\\%compression is positive now!!!
        &\epsilon_{\hat{\theta}\hat{r}}^{\mathrm{pt}}=\epsilon_{\hat{r}\hat{\theta}}^{\mathrm{pt}}=-\frac{1}{2} f \frac{\sigma_1}{E^{\prime}} \frac{\ell_{\mathrm{t}}}{c}\int_{-1}^{1} \Omega_r\text{d}r^{\prime}= -\frac{1}{2} f \frac{\sigma_1}{E^{\prime}} \frac{\ell_{\mathrm{t}}}{c}\left(\Omega_r(-1) -\int_{-1}^{1}r^{\prime} \frac{\partial \Omega_r}{\partial r^{\prime}} \text{d}r^{\prime}\right)\\
        & \epsilon_{\hat{r}\hat{r}}^{\mathrm{pt}}=0
    \end{aligned}
\end{equation}
In the global coordinate system of $x$, $y$:
\begin{equation}
    \begin{aligned}
    &\epsilon_{yy}^{\mathrm{pt}}=\sin{(\gamma-\theta)}(-2\epsilon_{\hat{\theta}\hat{r}}^{\mathrm{pt}}\cos{(\gamma-\theta)}+\epsilon_{\hat{\theta}\hat{\theta}}^{\mathrm{pt}}\sin{(\gamma-\theta)})\\
    &\epsilon_{yx}^{\mathrm{pt}}=\epsilon_{xy}^{\mathrm{pt}}=-\epsilon_{\hat{\theta}\hat{r}}^{\mathrm{pt}} \cos{2(\gamma-\theta)} +\dfrac{1}{2} \epsilon_{\hat{\theta}\hat{\theta}}^{\mathrm{pt}} \sin{2(\gamma-\theta)}\\
    &\epsilon_{xx}^{\mathrm{pt}}=\cos{(\gamma-\theta)}(\epsilon_{\hat{\theta}\hat{\theta}}^{\mathrm{pt}} \cos{(\gamma-\theta)} + 2 \epsilon_{\hat{\theta}\hat{r}}^{\mathrm{pt}}\sin{(\gamma-\theta)})
    \end{aligned}
\end{equation}
Note that the inelastic strains reported above are proportional to the dimensionless groups of material properties $f \tau_{\mathrm{y}}/E^{\prime}$ and loading conditions $\sigma_1/\tau_{\mathrm{y}}$. The normalized inelastic strain $(E^{\prime}/\tau_{\mathrm{y}})\epsilon^{\mathrm{p}}/f$ thus only scales with dimensionless stress ratio $\sigma_1/\tau_{\mathrm{y}}$.

\section{Results with different friction coefficients\label{ax:Results}}
\begin{figure}
    \centering
    \begin{tabular}{cc}
    \includegraphics[width=0.45 \linewidth]{ 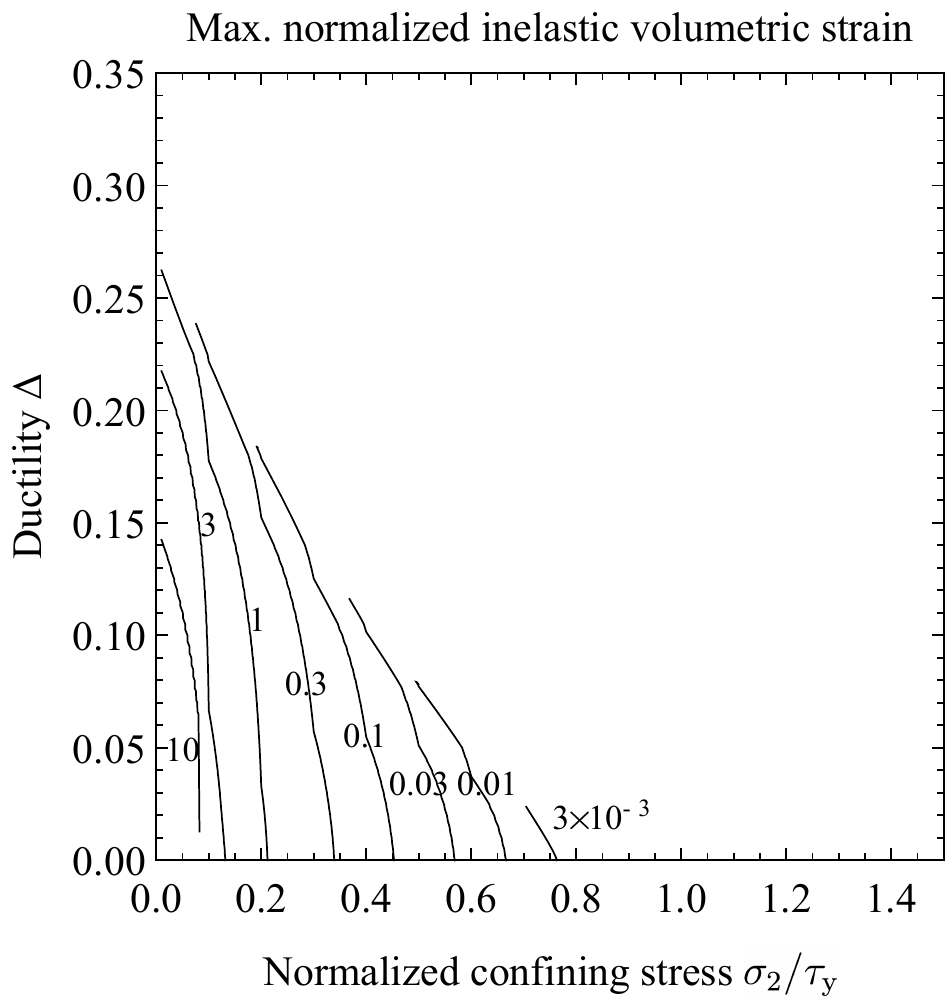}&
    \includegraphics[width=0.45 \linewidth]{ 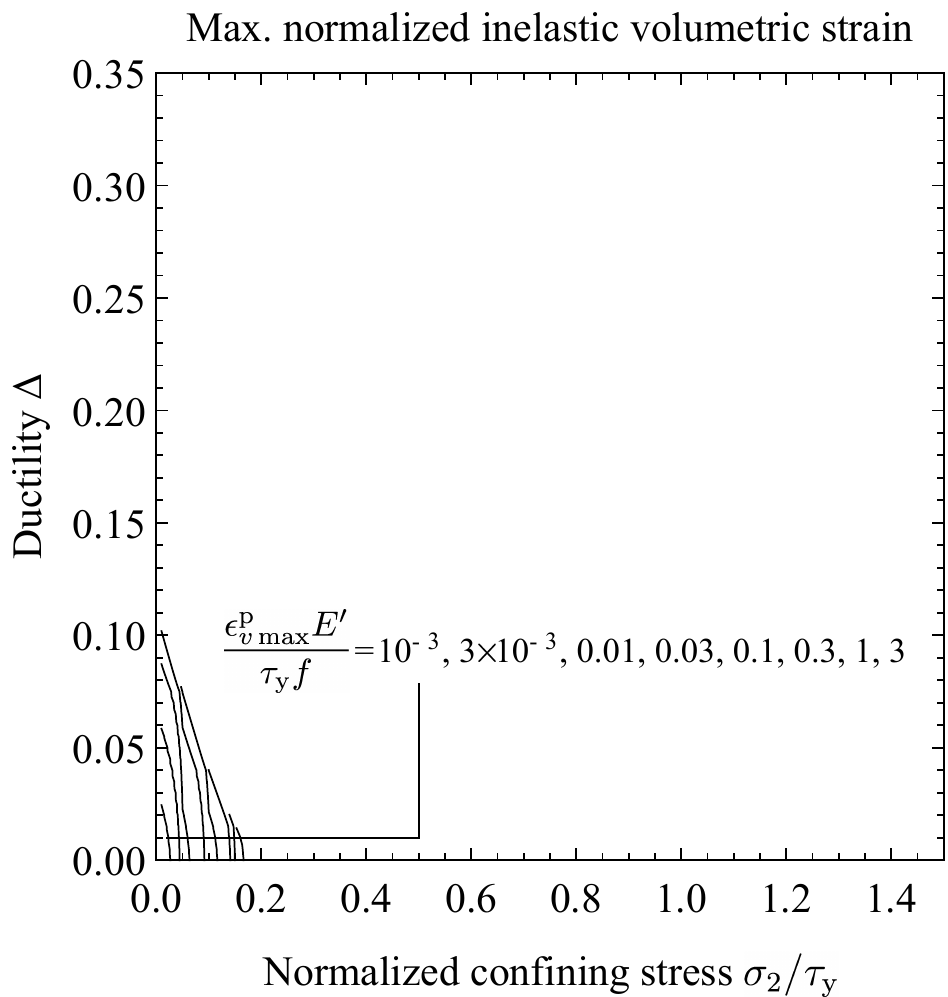}
    \\
    (a) $\mu=0.3$ & (b) $\mu=0.8$\\
    \end{tabular}
    \caption{Contours of constant normalized volumetric strains $\epsilon_{v\, \mathrm{max}}^{\mathrm{p}} (E^{\prime}/\tau_{\mathrm{y}})/f$ for the cases of $\gamma=\theta=\pi/4$ and $\tau_{\mathrm{c}}=0$ with different friction coefficients.
    }
    \label{fig:EvContour1}
\end{figure}

\begin{figure}
    \centering
    \begin{tabular}{cc}
    \includegraphics[height=0.45 \linewidth]{ 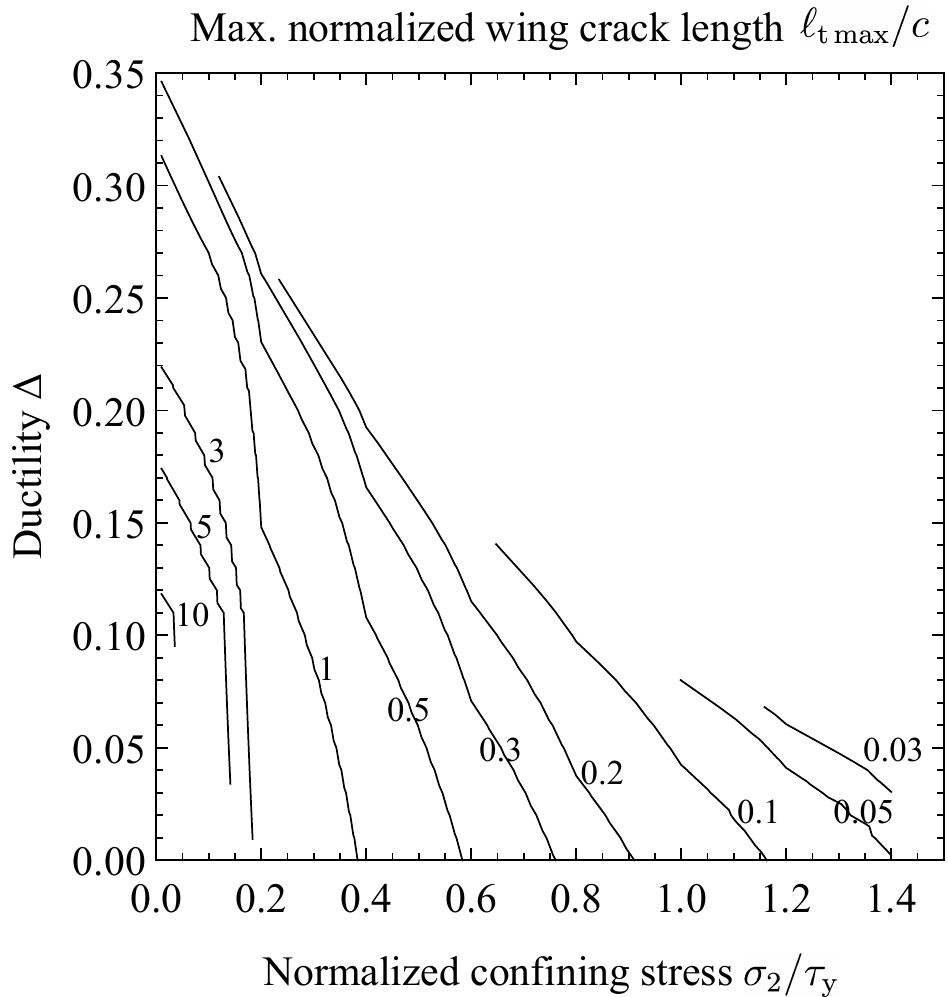}&
    \includegraphics[height=0.45 \linewidth]{ 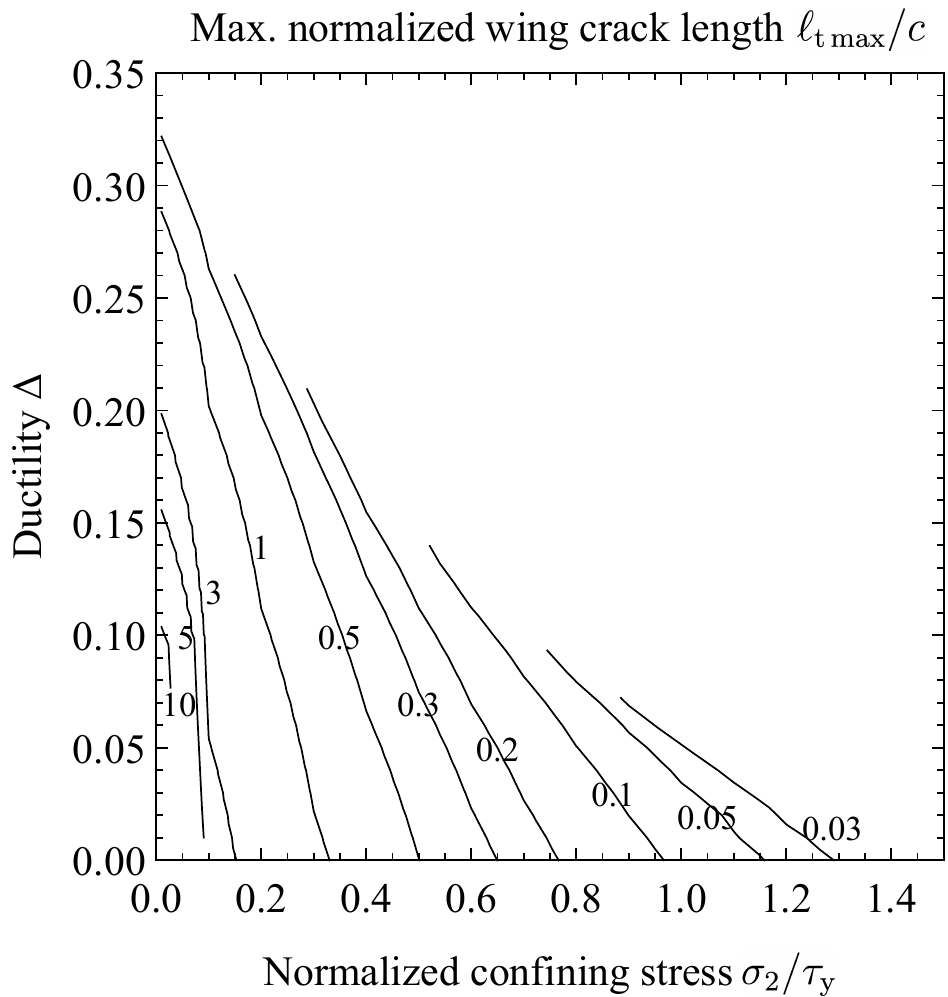}\\
    (a) $\mu=0, \tau_{\mathrm{c}}/\tau_{\mathrm{y}}=0$ & (b) $\mu=0.1, \tau_{\mathrm{c}}/\tau_{\mathrm{y}}=0$\\
    \includegraphics[height=0.45 \linewidth]{ 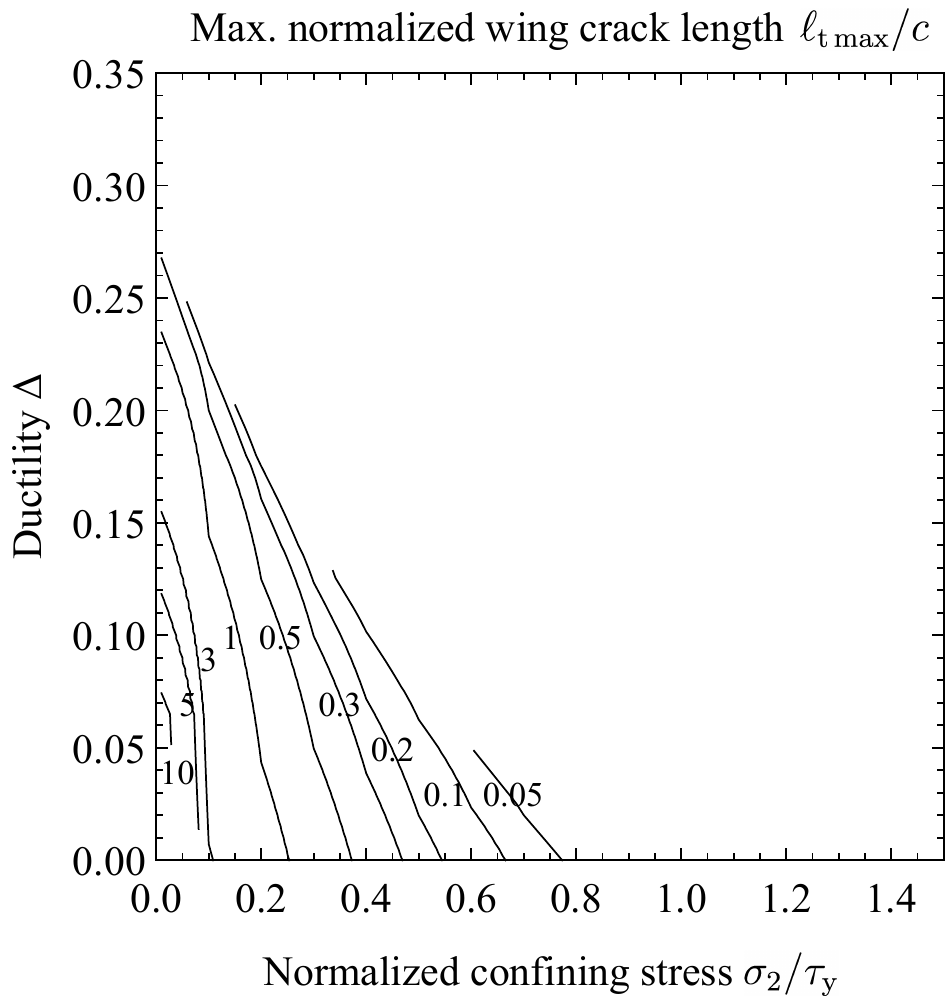} &
    \includegraphics[height=0.45 \linewidth]{ 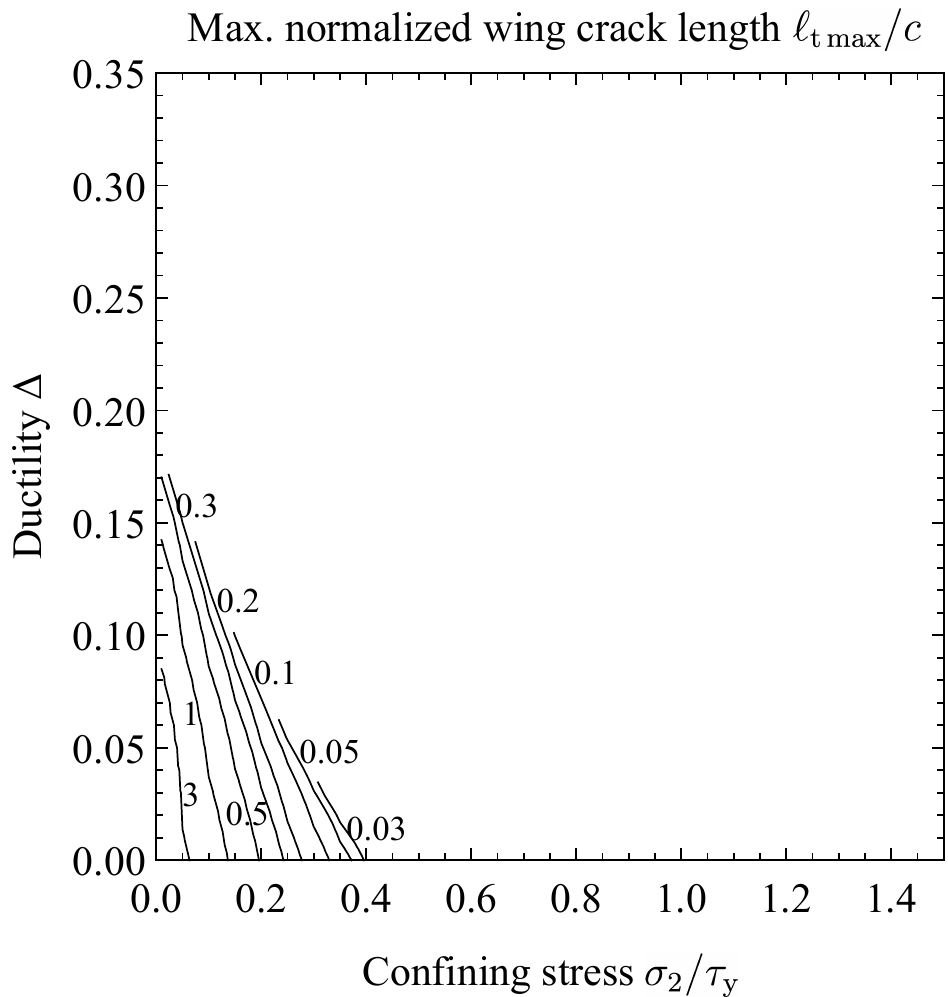}\\
    (c) $\mu=0.3, \tau_{\mathrm{c}}/\tau_{\mathrm{y}}=0$ & (d) $\mu=0.6, \tau_{\mathrm{c}}/\tau_{\mathrm{y}}=0$ \\
    \includegraphics[height=0.45 \linewidth]{ 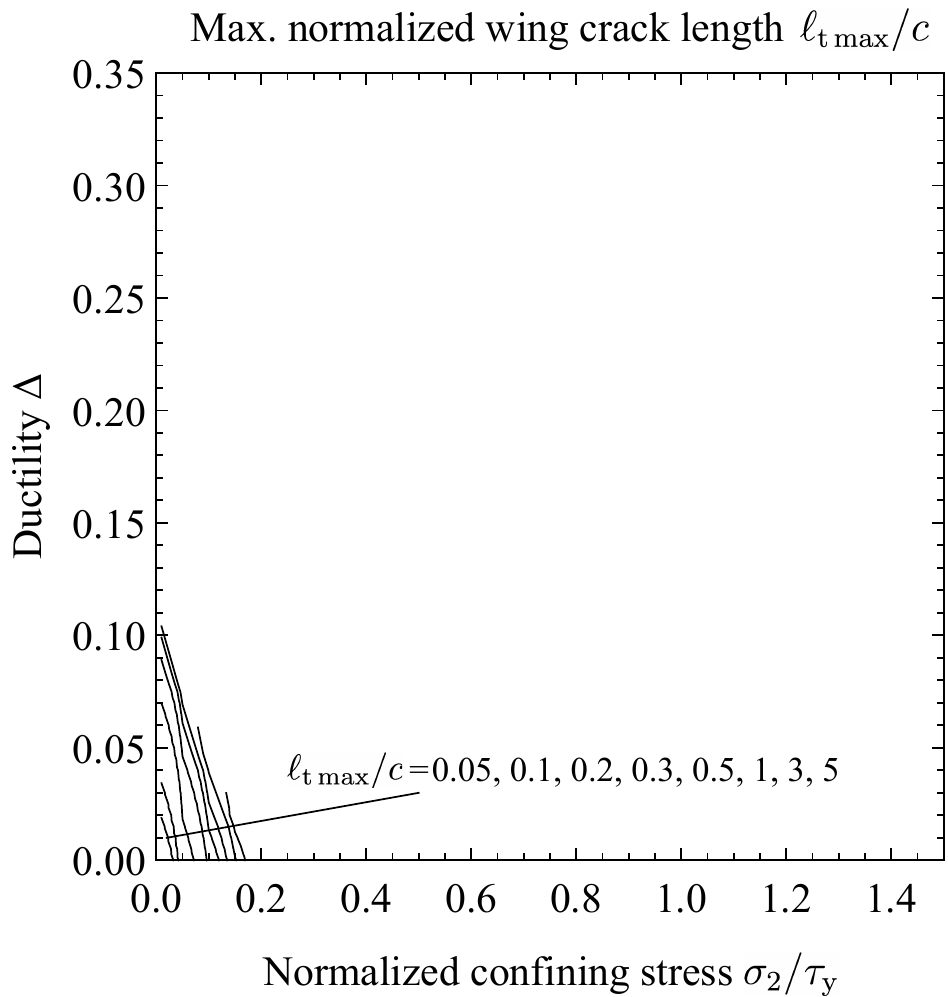} &
    \includegraphics[height=0.45 \linewidth]{ 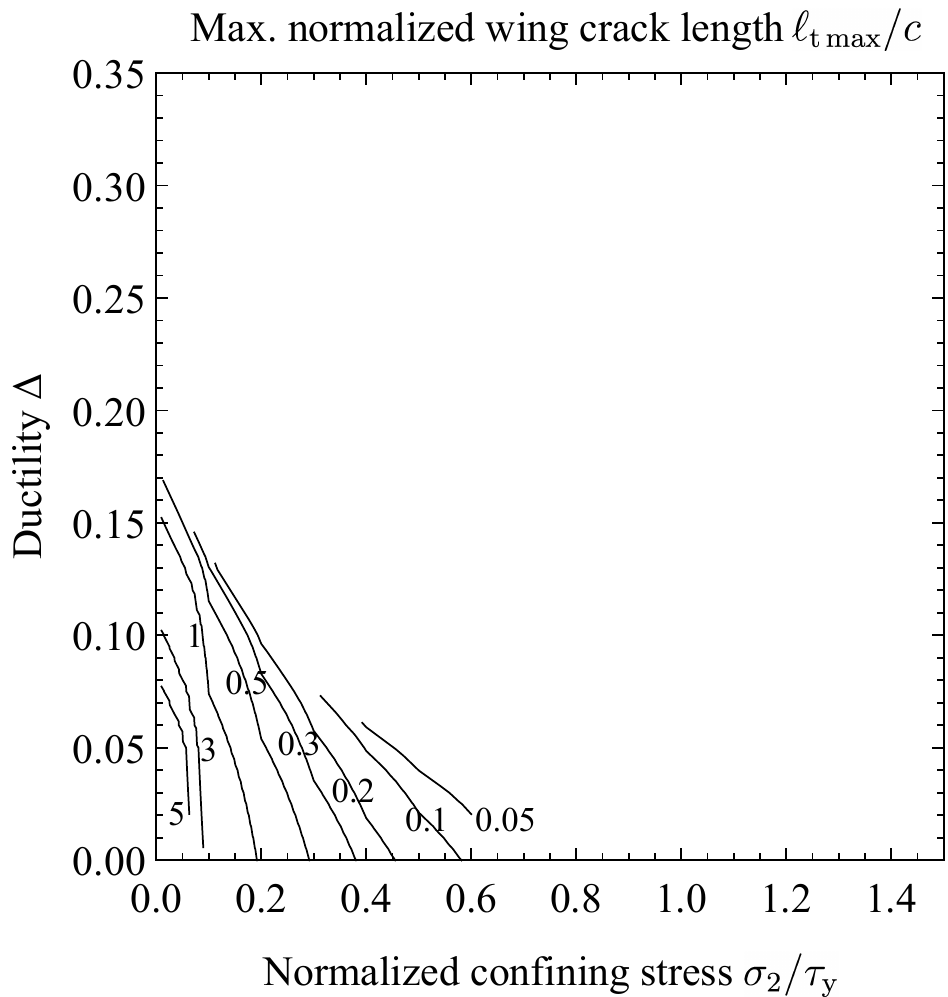}\\
    (e) $\mu=0.8, \tau_{\mathrm{c}}/\tau_{\mathrm{y}}=0$ & (f) $\mu=0, \tau_{\mathrm{c}}/\tau_{\mathrm{y}}=0.5$
    \end{tabular}
    \caption{Contours of the maximum wing crack length $\ell_{\mathrm{t}\, \mathrm{max}}$ for the cases of $\gamma=\theta=\pi/4$ with different friction coefficients and cohesion.
    }
    \label{fig:EvContour2}
\end{figure}

We show complimentary results with different friction coefficients in Fig.~\ref{fig:EvContour1}. These results together with those reported in Fig.~\ref{fig:EvContour} and Fig.~\ref{fig:mu0} are obtained without setting any constraints for the growth of the plastic zone and wing cracks. We show in Fig.~\ref{fig:EvContour2} the corresponding evolution of the maximum wing crack length for the cases reported in Fig.~\ref{fig:EvContour}, Fig.~\ref{fig:mu0} and Fig.~\ref{fig:EvContour1}.

%\end{linenumbers}% comment to remove line numbers

\end{document}